\documentclass[11pt,a4paper]{article}
\pdfoutput=1
\usepackage{jheppub, hyperref, cleveref}
\usepackage{amsfonts, amsmath, amssymb, latexsym}
\usepackage{graphicx, caption, subcaption}
\usepackage{comment}
\usepackage[normalem]{ulem}

\title{Measuring Parity Violation in the Stochastic Gravitational Wave Background with the LISA-Taiji network}

\author[a]{Giorgio Orlando}
\author[b]{\!\!, Mauro Pieroni}
\author[c,d]{\!\!, Angelo Ricciardone}

\affiliation[a]{Van Swinderen Institute for Particle Physics and Gravity, University of Groningen, Nijenborgh 4, 9747 AG Groningen, The Netherlands}
\affiliation[b]{Blackett Laboratory, Imperial College London, SW7 2AZ, UK}

\affiliation[c]{Dipartimento di Fisica e Astronomia ``G. Galilei",
	Universit\`a degli Studi di Padova, via Marzolo 8, I-35131 Padova, Italy}

\affiliation[d]{INFN, Sezione di Padova, via Marzolo 8, I-35131 Padova, Italy}

\emailAdd{g.orlando@rug.nl}
\emailAdd{m.pieroni@imperial.ac.uk}
\emailAdd{angelo.ricciardone@pd.infn.it}

\abstract{Parity violation is a powerful observable to distinguish a cosmological background of Gravitational Waves (GWs) from an astrophysical one. Planar single GW interferometers, both on ground and in space, are unable to measure the net circular polarization of an isotropic Stochastic Gravitational Wave Background (SGWB). In this paper, we explore the possibility of detecting circular polarization of an isotropic SGWB by cross-correlating two space-based detectors planned to be launched around 2034: LISA and Taiji. We compute the response of such a network to chirality and we perform a Fisher forecast analysis on the $I$ and $V$ Stokes parameters for the SGWB. We find that a clear measurement of chirality can be claimed  for a maximally chiral flat signal with amplitude  $h^2 \, \Omega_{\rm GW} \simeq 10^{-12}$ at the frequency scales of LISA and Taiji.}

\begin{document}
\begin{flushright}
	Imperial/TP/2020/MP/05
\end{flushright}
\maketitle

\section{Introduction}
\label{sec:intro}
The recent claim of the NANOGrav collaboration of a detection of a stochastic common-spectrum process from the 12.5-yr data set~\cite{Arzoumanian:2020vkk},
has pushed the attention towards the possibility to detect a cosmological Stochastic Gravitational Wave Background (SGWB), using current and future GW detectors. The gap in frequency between Pulsar Timing Array and ground-based detectors will be covered by space-based interferometers, which will work in the mHz regime.  For the next future, planned space-based detectors will be the Laser Interferometer Space Antenna (LISA)~\cite{Audley:2017drz} and Taiji~\cite{Guo:2018npi}, that will both consist of a constellation of three satellites forming a nearly equilateral triangle with 2.5 and 3 million km arm lengths respectively. Since such detectors will be probably flying on the same time, it is interesting to study the capability of such network of extracting information on the SGWB. 

Recently, a lot of effort has been dedicated to the development of techniques and tools to characterize the SGWB using a triangular detector like LISA~\cite{Caprini:2018mtu, Caprini:2015zlo, Bartolo:2016ami,Caprini:2019egz,Auclair:2019wcv, Karnesis:2019mph, Flauger:2020qyi, Barausse:2020rsu, Pieroni:2020rob, Caprini:2019pxz,Contaldi:2020rht} or correlating different experiments~\cite{Campeti:2020xwn}. For an isotropic GW background, single planar detectors are limited by their symmetry in detecting crucial observables, like chirality, which would be very important in the process of characterization and disentanglement of a cosmological signal from an astrophysical one~\cite{Seto:2007tn, Seto:2008sr, Smith:2016jqs}. The reason is that a planar interferometer responds in the same way to a left-handed GW arriving perpendicular to the plane of the detector and to a right-handed GW of the same amplitude coming from the opposite direction. On the contrary, if the SGWB is anisotropic then a single planar detector (like LISA) becomes sensitive to chiral signals. For a recent estimate of the sensitivity of LISA to circular polarization using a dipolar anisotropy see~\cite{Domcke:2019zls}.

It has recently been pointed out~\cite{Seto:2020zxw} that in the case of an isotropic background, by cross-correlating the data streams of the LISA and Taiji, we can measure the chirality of the SGWB, described by the Stokes parameter $V$, which characterizes the asymmetry between the amplitudes of the left- and right-handed polarized waves. Such a measurement is important to test many early universe theories which predict a sizable degree of parity violation, like models where the inflaton is coupled with gauge-fields or when Chern-Simons couplings~\footnote{Notably this coupling also provides a very efficient channel for gauge preheating~\cite{Adshead:2015pva, Adshead:2016iae, Cuissa:2018oiw} which typically produces a sizable GW signal peaked at high frequencies~\cite{Adshead:2018doq, Adshead:2019lbr, Adshead:2019igv}.} are present~\cite{Lue:1998mq,Jackiw:2003pm, Alexander:2004wk,Contaldi:2008yz,Takahashi:2009wc,Satoh:2010ep,Sorbo:2011rz, Barnaby:2011qe,Maleknejad:2011jw, Shiraishi:2013kxa,Creminelli:2014wna,Maleknejad:2016qjz,Dimastrogiovanni:2016fuu, Domcke:2016bkh, Bartolo:2017szm,Bartolo:2018elp,Domcke:2017fix,Domcke:2018rvv,McDonough:2018xzh,Qiao:2019hkz,Mirzagholi:2020irt,Watanabe:2020ctz, Bartolo:2020gsh,Bordin:2020eui, Almeida:2020kaq, Ozsoy:2020ccy}.

In this paper, as original contribution, we perform a detailed Fisher forecast analysis for the $I$ and $V$ Stokes parameters, with the aim of quantifying how well we can constrain chirality with the LISA-Taiji network. As expected, we find that the self-correlations between channels of the same detector only contribute to the measurement of $I$, while the cross-correlations of different detector channels contribute also to $V$. By taking into account Taiji orbit specifications given in recent papers~\cite{Ruan:2020smc, Omiya:2020fvw, Seto:2020mfd},  the most up-to-date LISA instrument specifications~\cite{PhysRevLett.120.061101, ldcdoc, Flauger:2020qyi}, and assuming same mission duration and scan strategy for LISA and Taiji, we derive the expected $1$ and $2\sigma$ error bars around the best fit value in the $I$-$V$ plane. We analyze both the case of an unpolarized background and of a maximally chiral background for different signal amplitudes. In the latter case, we find that the estimated error on $V$ is roughly between 1 and 2 orders of magnitude larger than the estimated error on $I$. From our results, we can state that a clear measurement of chirality can be claimed  for a maximally chiral flat signal with amplitude  $h^2 \, \Omega_{\rm GW} \simeq 10^{-12}$ at the frequency scales of LISA and Taiji.

Throughout the paper,  we also derive the (cross-correlated) detector responses to a  SGWB, working in the TDI 1.5 AET channel basis~\cite{Hogan:2001jn, Adams:2010vc} and compute the Signal-to-Noise Ratio (SNR) for the detection of chirality, comparing our results with~\cite{Seto:2020zxw}. 
\\
The structure of the paper is the following: in \cref{sec:network} we describe the orbit configuration of the two interferometers and we compute the  response functions for measuring the net circular polarization of the SGWB with the LISA-Taiji network. In \cref{sec:SNR} we compute the corresponding SNR for chiral SGWB. In \cref{sec:fisher} we perform a Fisher forecast analysis on the Intensity ($I$) and Circular Polarization ($V$) parameters. In \cref{sec:conclusions} we draw our conclusions. Finally, in \cref{sec:appendix} we report some useful formulae for the computation of the response functions and noise spectra, mainly taken from~\cite{Flauger:2020qyi}.

\section{LISA-Taiji detector network and detection of chirality}
\label{sec:network}

\begin{figure}
	\centering
	\begin{subfigure}[h]{.49 \textwidth}
		\centering
		\vspace{0pt}
		\includegraphics[width=\linewidth]{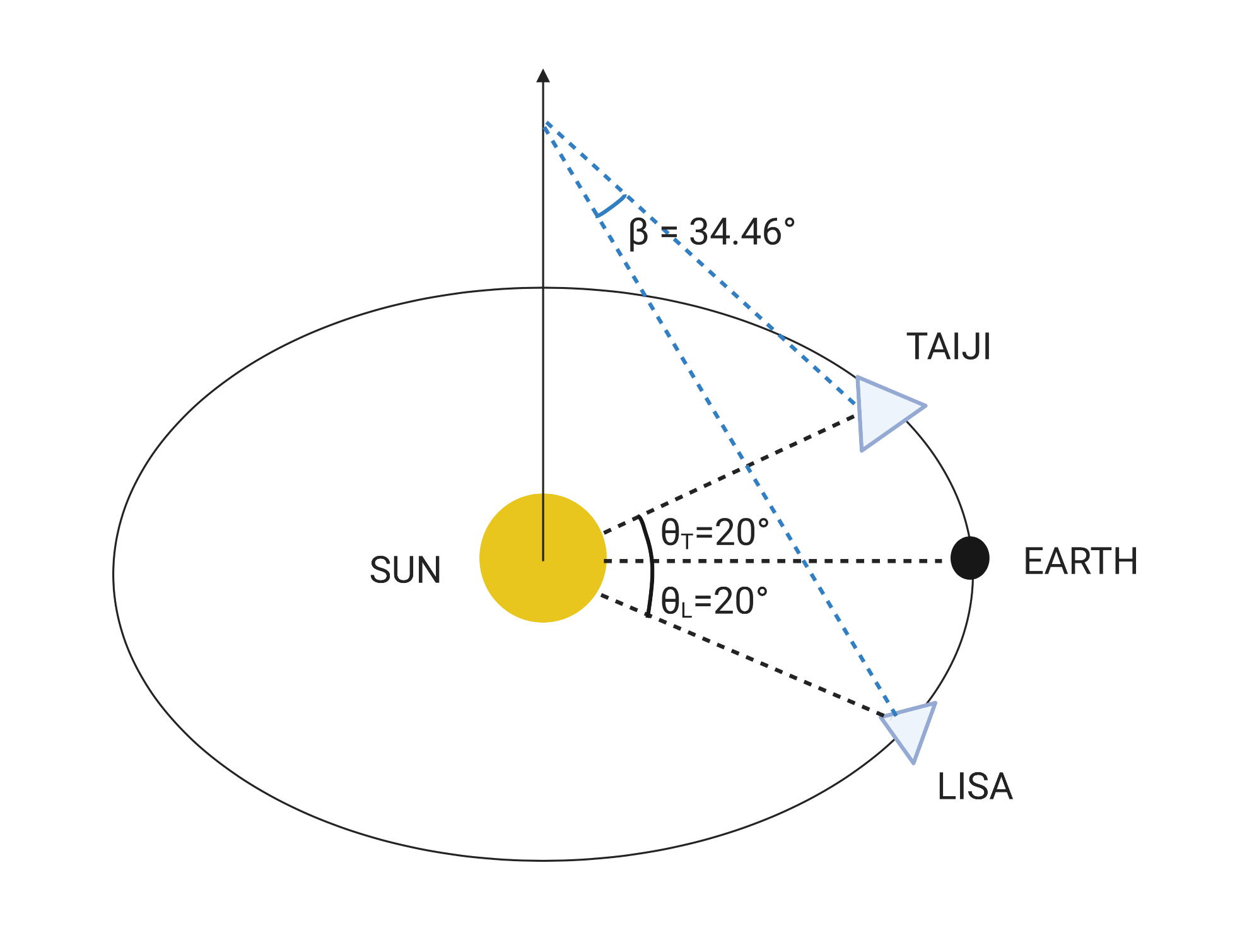}
	\end{subfigure}
	\begin{subfigure}[h]{.49\textwidth}
		\centering
		\vspace{0pt}
		\includegraphics[width=\linewidth]{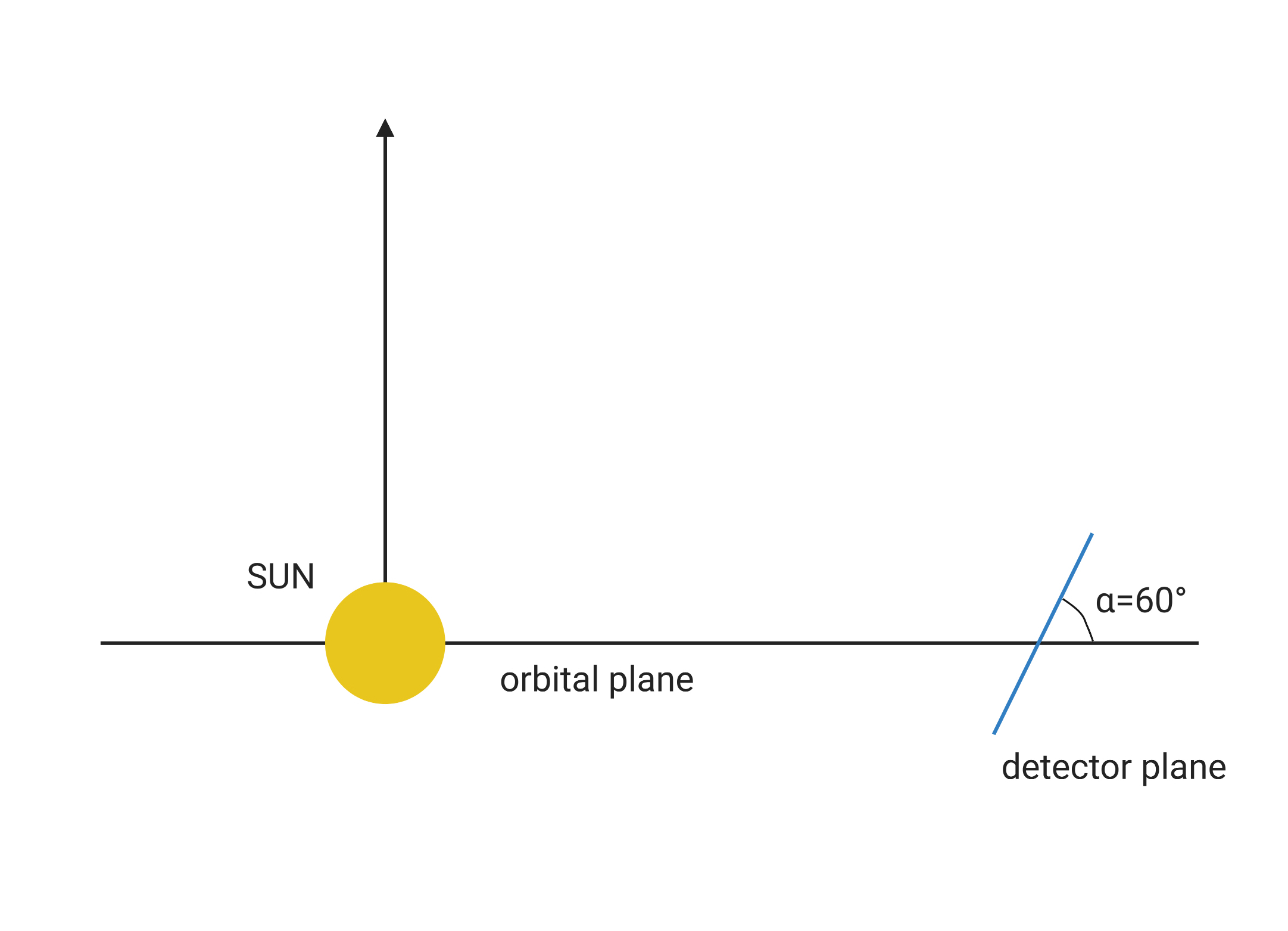}
	\end{subfigure}
	
\caption{\it Plot of the Lisa-Taiji network configuration. On the left, we show the global geometry of the LISA-Taiji network, with LISA having an orbital phase angle  $\theta_L = 20^o$ behind the Earth, and Taiji an orbital phase angle $\theta_T = 20^o$ ahead the Earth. The angular separation between the two detectors is $\beta = 34.46^o$. On the right, we show the plane of the two detectors forming an angle of $\alpha_L = \alpha_T = 60^o$ with the ecliptic plane. \label{fig:network}}

\end{figure}

We begin this section with a brief description of the  LISA-Taiji detector network (see, e.g.,~\cite{Ruan:2020smc} and refs. therein for more details). LISA is expected to move in an heliocentric (elliptical) orbit staying $20^o$ behind the Earth, with its three spacecrafts forming an equilateral triangle with the arm length $L_{\rm Lisa} = 2.5 \times 10^6 \, \mbox{km}$. The Taiji detector will share the same geometry and path of LISA, but staying $20^o$ ahead the Earth and having a larger arm length of $L_{\rm Taiji} = 3 \times 10^6 \, \mbox{km}$. For both LISA and Taiji, the planes of the three spacecrafts will be inclined of $60^o$ with respect to the ecliptic plane. Moreover, the detectors have an internal spinning motion (the so-called cartwheel rotation) with a period of 1 year. See Fig.~\ref{fig:network} for a qualitative representation of the LISA-Taiji network configuration. Since the Earth orbital eccentricity (and thus LISA and Taiji) $e$ is small, \emph{i.e.} $e \simeq 0.0167$, throughout this paper we neglect the eccentricity and assume circular orbits.

\begin{figure}
	\centering
	\begin{subfigure}[t]{.49 \textwidth}
		\centering
		\vspace{0pt}
		\includegraphics[width=\linewidth]{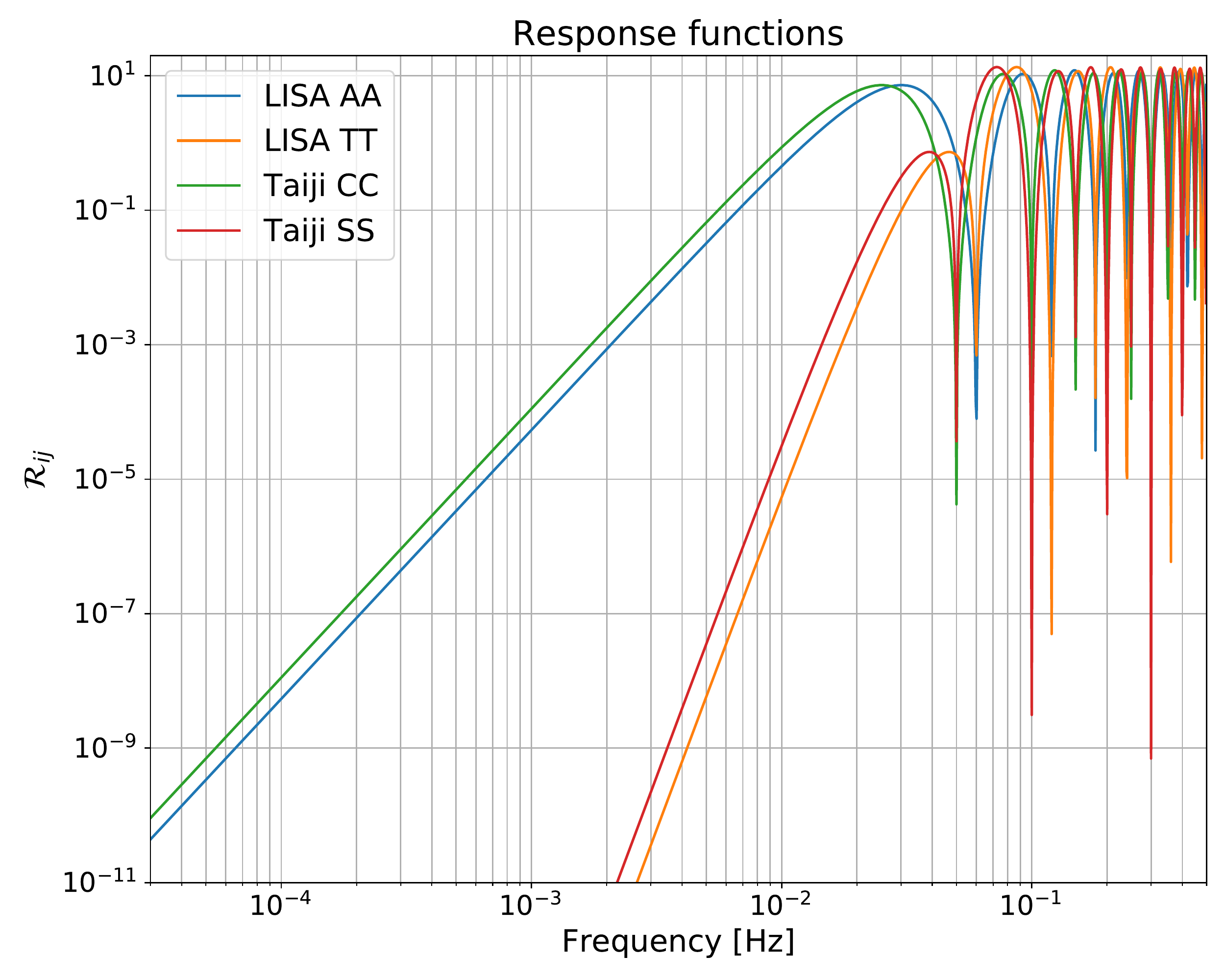}
		\caption{}\label{fig:fig_a}
	\end{subfigure}
	\begin{subfigure}[t]{.49\textwidth}
		\centering
		\vspace{0pt}
		\includegraphics[width=\linewidth]{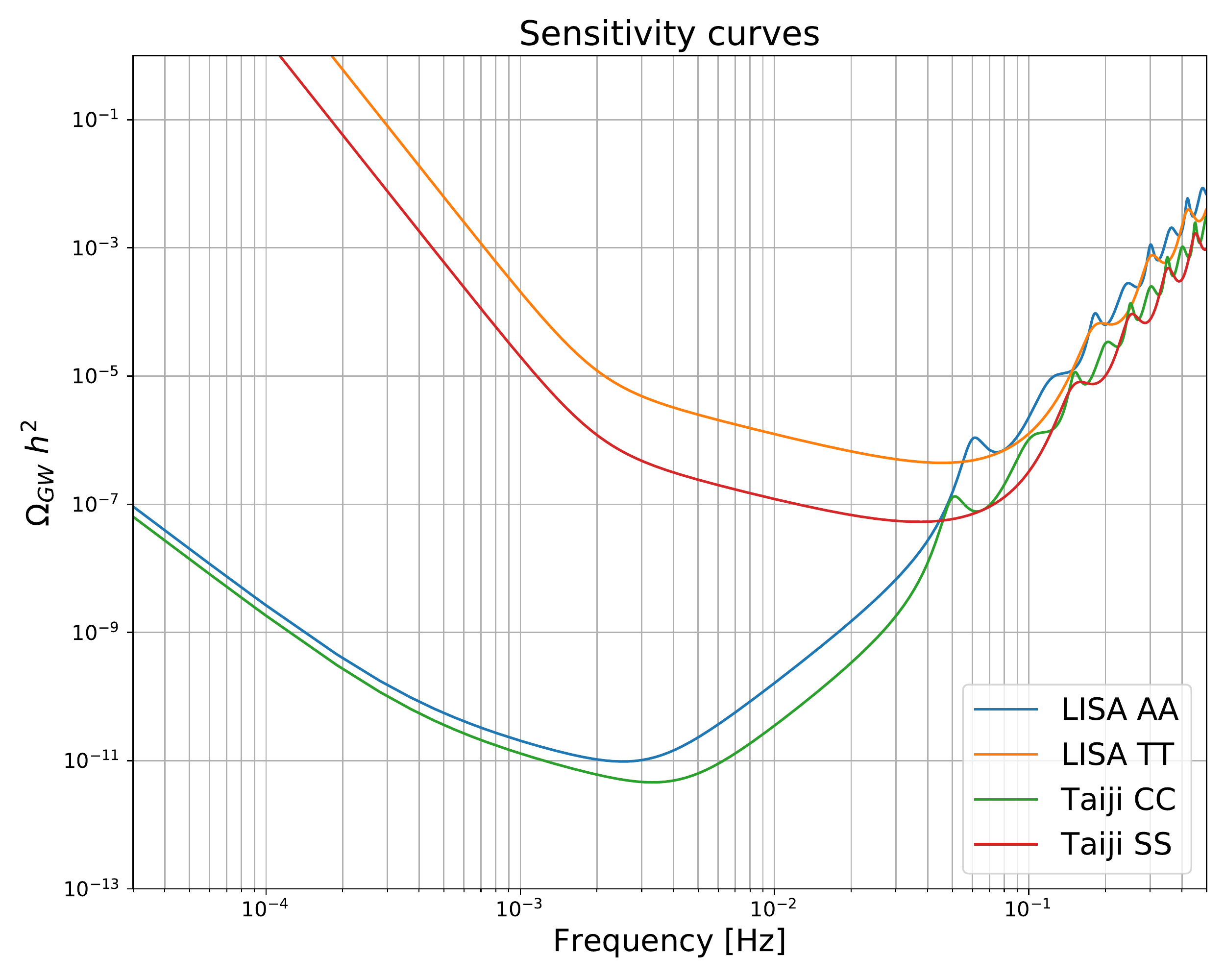}
		\caption{}\label{fig:fig_b}
	\end{subfigure}

	\medskip
	
	\begin{subfigure}[t]{.49\textwidth}
		\centering
		\includegraphics[width=\linewidth]{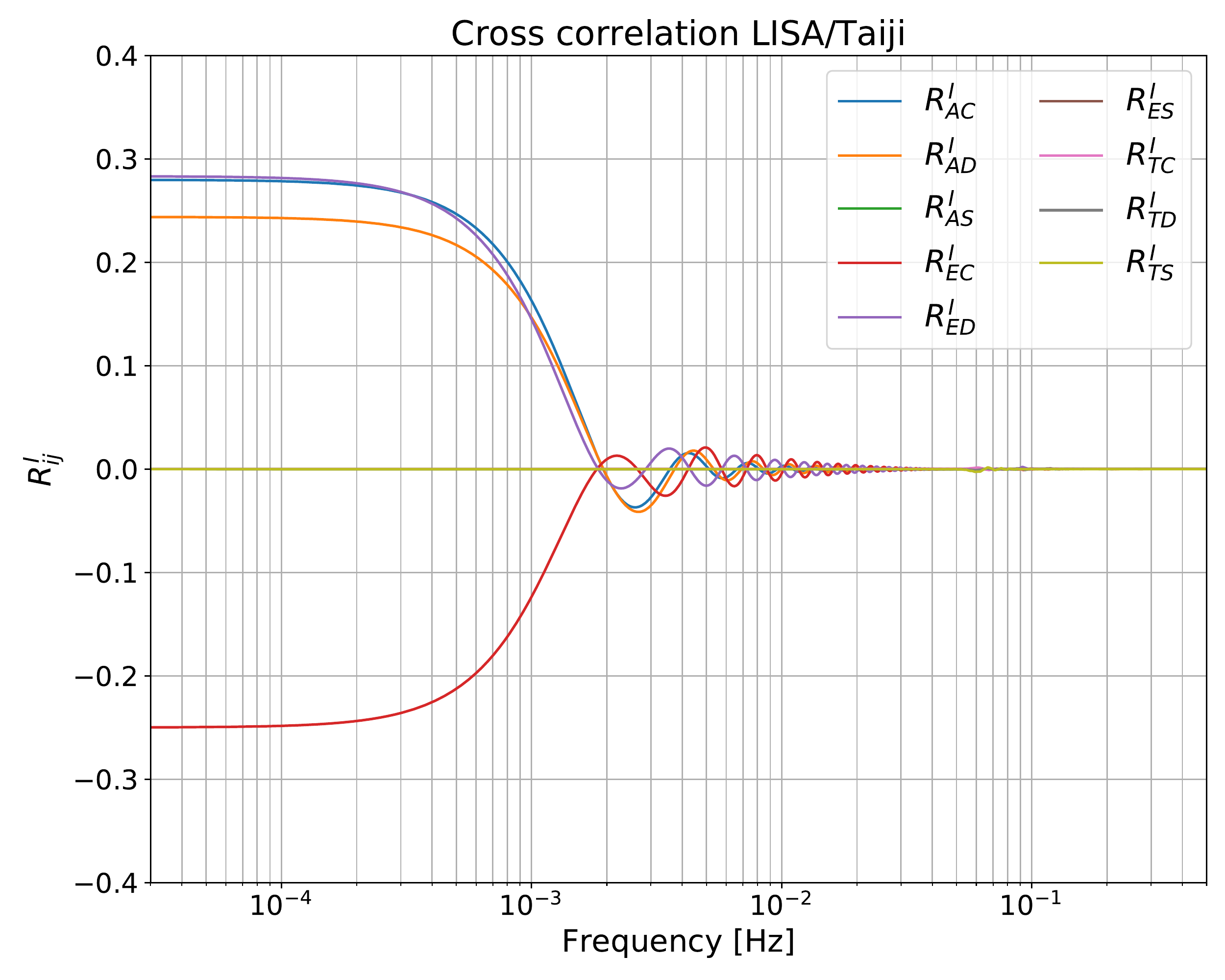}
		\caption{}\label{fig:fig_c}
	\end{subfigure}
	\begin{subfigure}[t]{.49\textwidth}
		\centering
		\includegraphics[width=\linewidth]{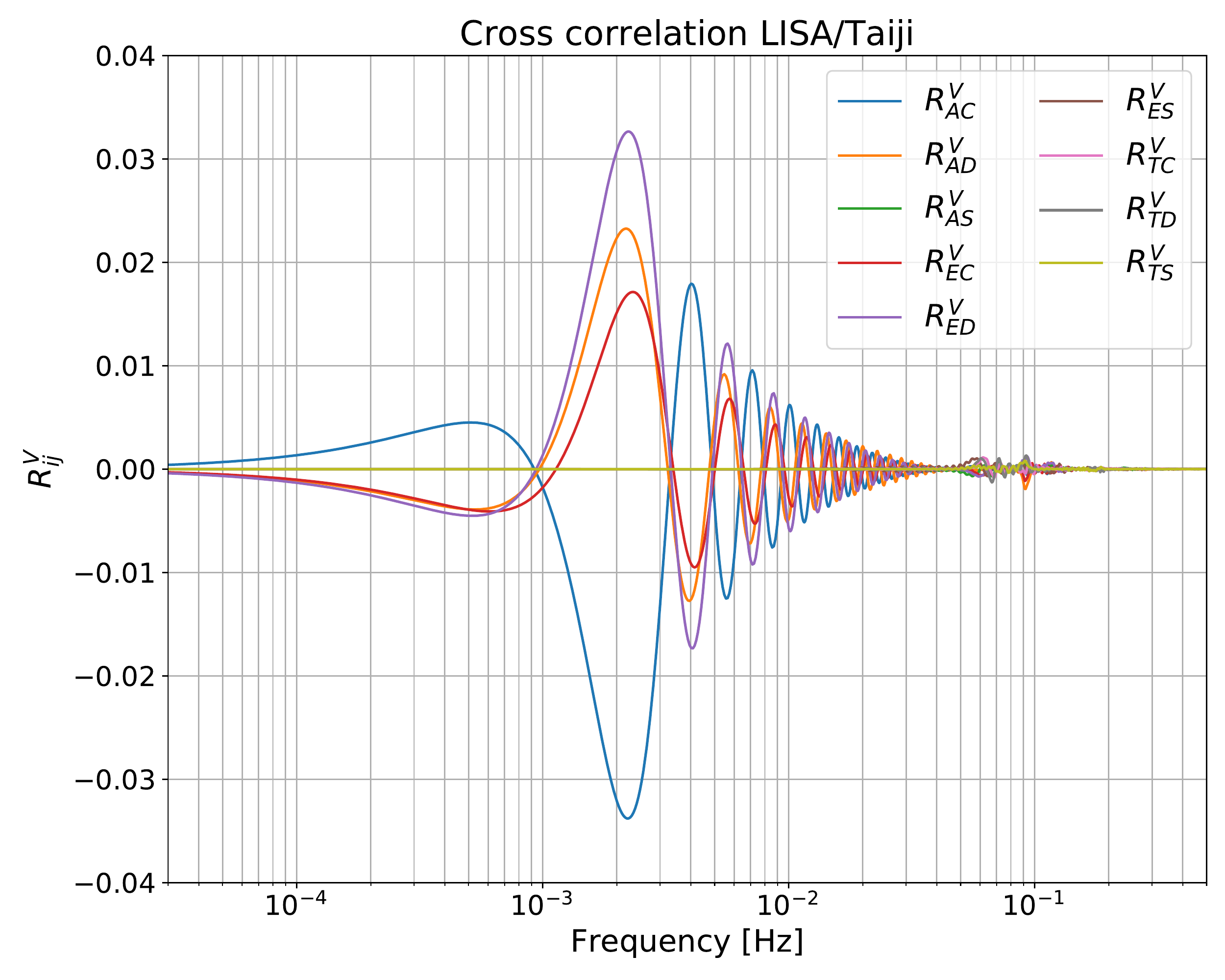}
		\caption{}\label{fig:fig_d}
	\end{subfigure}

	\caption{\it Response functions and sensitivity curves for the Lisa-Taiji detector network. In panel (a) we show the single detector $\mathcal{R}_{ij}(f) = \mathcal{R}^L_{ij}(f) + \mathcal{R}^{R}_{ij}(f)$ response functions. In panel (b) we show the sensitivity curves for LISA and Taiji noise-diagonal channels. In panel (c) we show the cross-correlation $\tilde{R}^{I}_{ij}(f)$  response function. In panel (d) we show the cross-correlation $\tilde{R}^{V}_{ij}(f)$  response function. \label{fig:response}}

\end{figure}

Each detector consists of three interferometers (data channels) which simultaneously measure the differential Doppler frequency shifts induced by GWs passing between the test masses. In general, the data stream (in time domain) measured in any of these channels (labeled by the index $i$) can be modeled as
\begin{equation}
	d_i(t)  = s_i(t) + n_i(t) \; , 
\end{equation}
where $n_i(t)$ is the noise of the instrument and $s_i(t)$ is some residual signal. Typically, it is more convenient to work in frequency domain, which can be achieved through a finite-time Fourier transform:
\begin{equation}
	\label{eq:Fourier}
	\tilde d_i(f) = \int_{-T/2}^{T/2} \,dt \, e^{2 \pi i  f t} \, d_i (t) \, .
\end{equation}
where  $T$ is the observation time.
 In the following we assume the signal and the noise to be Gaussian (with zero mean) and uncorrelated (\emph{i.e.}, $\langle \tilde{s}_i(f)   \tilde{n}_j(f') \rangle = 0$). In this case, the information is contained in the signal and noise power spectra which, for stationary signals, can be expressed as~\footnote{We have already replaced the finite-time delta function with an ordinary Dirac delta function, assuming infinite observation time $T$.} 
\begin{equation}
	\begin{aligned}
		\langle \tilde s_i(f) \tilde s_j^*(f') \rangle = \frac{1}{2} S_{ij}(f) \, \delta(f - f')  \, , \\
		\langle \tilde n_i(f) \tilde n_j^*(f') \rangle = \frac{1}{2} N_{ij}(f) \, \delta(f - f')  \, , \\
	\end{aligned}
\end{equation}
where the quantity $S_{ij}(f), N_{ij}(f)$ are respectively the so-called ``one-sided" signal and noise power spectra, obeying to $S_{ij}(-f) = S_{ij}(f)$, $N_{ij}(-f) = N_{ij}(f)$. By taking the Fourier transform of~\cref{eq:two_point_dfreq2_app2}, it is possible to show that $S_{ij}$ can be expressed as:
\begin{eqnarray}
	\label{eq:power_spectrum_resp}
S_{ij}(f) = \sum_{\lambda} \mathcal{R}^\lambda_{ij} (f) P_{\lambda}(f) =  \sum_{\lambda} P_{\lambda}(f)  \left[   (2\pi k L_i) (2\pi k L_j) \, W(k L_i) \, W^*(k L_j) \tilde{R}^\lambda_{ij} (f)  + h.c. \right] \; , \nonumber\\
\end{eqnarray}
where $\lambda = L/R$ identifies left and right-handed polarizations (see~\cref{sec:formulae}), $L_i, L_j$ are the detector armlengths, $\mathcal{R}^\lambda_{ij}$ is the so-called detector response function and  $P_{\lambda}$ is the GW power spectrum defined as
\begin{equation}
	\label{eq:spectrum}
	\langle \tilde{h}_{\lambda} (\mathbf{k}_1) \tilde{h}^*_{\lambda'} (-\mathbf{k}_2)\rangle 
	= \delta^{(3)}(\mathbf{k}_1 +\mathbf{k}_2) \frac{ P_{\lambda}(k_1) }{4 \pi k_1^2}\delta_{\lambda\lambda'} \; ,  \hspace{1cm} \langle \tilde{h}_{\lambda} (\mathbf{k}_1) \tilde{h}_{\lambda'} (\mathbf{k}_2)\rangle 
	=\ 0 \; .
\end{equation}
For the scope of this work, it is more useful to switch to the so-called $I$ and $V$ Stokes parameters defined as
\begin{equation}
	\label{eq:stokes}
	I = P_R+ P_L \, , \qquad \qquad	V = P_R -P_L  \, .
\end{equation}
Here $I$ defines the overall total amplitude of gravitational waves, while $V$ defines its circular polarization, \emph{i.e.} the asymmetry between the L and R-handed power spectra. The latter could be induced by mechanisms of parity violation in the Universe, thus representing a clear channel for looking to parity breaking signatures in the gravitational interaction. Clearly, it is possible to express $S_{ij}(f) $ in terms of $I$ and $V$ as
\begin{equation}
	\label{eq:two_point_dfreq2} 
	S_{ij}(f)   =     I (f) \mathcal{R}^{I}_{ij}(f)  +V(f) \mathcal{R}^{V}_{ij}(f)  \; , 
\end{equation}
where we have introduced the $I$ and $V$ detector responses as
\begin{equation}
	\label{eq:resp_stokes}
	\mathcal{R}^{I}_{ij}(f) = \frac{ \mathcal{R}_{ij}^{R}(f)+ \mathcal{R}_{ij}^L(f)  }{2}\, , \quad \quad \quad\quad \mathcal{R}^{V}_{ij}(f) = \frac{ \mathcal{R}_{ij}^{R}(f)- \mathcal{R}_{ij}^{L}(f)  }{2}\, .
\end{equation}
Before considering cross-correlations between different detectors, let us focus on the case where $i$ and $j$ denote two channels of the same detector. In this case, $\mathcal{R}^{I}_{ij}(f) $ matches with the function $\mathcal{R}_{ij}(f) $ of~\cite{Flauger:2020qyi}. A plot of this quantity for both LISA and Taiji is shown in~\cref{fig:fig_a}. Since both LISA and Taiji are planar detectors, they are insensitive to parity violations~\cite{Seto:2008sr, Thorne:2017jft}. This can be shown by checking that in the two cases we have $\mathcal{R}^L_{ij} = \mathcal{R}^R_{ij}$ and thus $\mathcal{R}^{V}_{ij}(f) = 0$. We introduce GW spectral energy density using
\begin{equation}
h^2	\Omega^{\lambda}_{\rm GW}(f) = \frac{4 \pi^2 f^3}{3 (H_0/h)^2} P_\lambda(f) \; , \qquad h^2	\Omega_{\rm GW}(f) = \frac{4 \pi^2 f^3}{3 (H_0/h)^2} I(f) \, ,
\end{equation}
where $H_0/h = 3.24 \times 10^{-18} 1/\text{s}$ is the value of the Hubble parameter at present time ($h$ is its dimensionless parameter normalization).

Both for LISA and Taiji it is possible to interpret the three output channels as data measured by three interferometers which share theirs arms. As a consequence, the noises in the three channels are correlated~\footnote{For the expressions of $N_{ \text{XX} }$ and $N_{ \text{XY} }$ for LISA and Taiji see~\cref{app:self_noise} and~\cref{app:cross_noise}.}.  Assuming that the noise spectra for all links are identical, it is possible to exploit the internal symmetry of each detector and build three noise-diagonal data combinations~\cite{Prince:2002hp}, \emph{i.e.} $N_{ij} =0 $ for all $i \neq j$ (see~\cref{app:noise} for more details). For LISA these data combinations are typically dubbed A, E, T  and, throughout this paper, the analogous combinations for Taiji will be dubbed C, D, S. In the following we will always proceed using these bases. In the LISA noise diagonal basis we have
\begin{equation}
	\hspace{-0.5cm}
	\begin{aligned}
		\label{eq:AA_noise}
		N_{ \text{AA} }(f, A, P)  & = N_{ \text{EE} }(f, A, P) =   \\
		& =  8 \left(\frac{2 \pi fL} {c}\right)^2 \sin^2\left(\frac{2 \pi f L}{c}\right) \left\{ 4 \left[1 +\cos \left(\frac{2 \pi f L}{c} \right) + \cos^2 \left(\frac{2 \pi f L}{c} \right)\right] \times \right. \\
		& \hspace{.5cm}\left. \times \frac{A^2}{L^2} \, \frac{\mbox{fm}^2}{\mbox{s}^4 \, \mbox{Hz}} \left[ 1 + \left(\frac{0.4 \mbox{mHz}}{f}\right)^2 \right] \left[ 1 + \left(\frac{f}{8 \mbox{mHz}}\right)^4 \right] \left(\frac{1}{2 \pi f} \right)^4 \;  + \right. \\ 
		& \hspace{1cm} \left. + \left[ 2 +\cos \left(\frac{2 \pi f L}{c} \right) \right]  \times   \,\frac{P^2}{L^2} \, \frac{\mbox{pm}^2}{\mbox{Hz}} \left[ 1 + \left(\frac{2 \mbox{mHz}}{f}\right)^4 \right]  \right\}  \; , \\
	\end{aligned}
\end{equation}
where $A$ and $P$ are amplitude of the acceleration and of the IMS noise~\footnote{See \cref{app:noise} for details.}, and
\begin{equation}
	\begin{aligned}
		\label{eq:TT_noise}
		N_{ \text{TT} }(f, A, P) & =  16 \left(\frac{2 \pi fL} {c}\right)^2 \sin^2\left(\frac{2 \pi f L}{c}\right) \left\{ 2 \left[ 1 - \cos \left(\frac{2 \pi f L}{c} \right) \right]^2  \times \right.  \\
		& \hspace{.5cm}\left. \times \frac{A^2}{L^2} \, \frac{\mbox{fm}^2}{\mbox{s}^4 \, \mbox{Hz}} \left[ 1 + \left(\frac{0.4 \mbox{mHz}}{f}\right)^2 \right] \left[ 1 + \left(\frac{f}{8 \mbox{mHz}}\right)^4 \right] \left(\frac{1}{2 \pi f} \right)^4 \;  + \right. \\ 
		& \hspace{2cm} \left. + \left[ 1 - \cos \left(\frac{2 \pi f L}{c} \right) \right]  \times   \,\frac{P^2}{L^2} \, \frac{\mbox{pm}^2}{\mbox{Hz}} \left[ 1 + \left(\frac{2 \mbox{mHz}}{f}\right)^4 \right]  \right\}  \; .
	\end{aligned}
\end{equation}
Identical expressions can be used for the CC/DD and SS Taiji noise power spectra. For the two detectors we respectively take $A=3$, $P=15$ and $A=3$, $P=8$~\cite{Ruan:2020smc}. In order to directly compare the injected signal with the noise, it is customary to define the strain sensitivities (see, e.g,~\cite{Flauger:2020qyi}) as
\begin{equation}
		S_{n,ij}(f) = \frac{N_{ij}(f)}{\sum_{\lambda} \mathcal{R}^{\lambda}_{ij}(f)} \, ,
\end{equation}
which can be expressed in $\Omega_{\rm GW}$ units as:
\begin{equation} \label{eq:Omega_units}
	h^2 \Omega_{n, ij}(f) = \frac{4 \pi^2 f^3}{3 (H_0/h)^2} S_{n,ij}(f) \equiv \frac{N^{\Omega}_{ij}(f)  }{\sum_\lambda \mathcal{R}^{\lambda}_{ij}(f)}\, .
\end{equation}
A plot of the sensitivities for the different data combinations for both LISA and Taiji (AA is equal to EE and CC is equal to DD) is shown in~\cref{fig:fig_b}.

So far, we have assumed that the two channels $i$, $j$ belong to a single detector. However, a lot of extra information can be unveiled by cross-correlating the signals of LISA and Taiji. 
In particular, since the LISA/Taiji network is not planar, by cross-correlating data streams from the two detectors it is possible to get information on $V$. In~\cref{fig:fig_c}, ~\cref{fig:fig_d} we respectively show~\footnote{Notice that these are the $\tilde{R}^\lambda_{ij} $ and not the $\mathcal{R}^\lambda_{ij}$ of~\cref{eq:power_spectrum_resp}.} $\tilde{R}^{I}_{ij} (k)$, $\tilde{R}^{V}_{ij} (k)$. Notice that the maximum of $\tilde{R}^{V}_{ij} (k)$ is roughly an order of magnitude smaller than $\tilde{R}^{I}_{ij} (k)$ for the same combination. Despite this suppression, the non-zero value of $\tilde{R}^{V}_{ij} (k)$ in this combination constitutes a remarkable improvement with respect to self-correlations of planar detectors where this quantity was shown to be identically vanishing. This open up a new channel for searching the chirality of gravitational waves with networks of space-based interferometers.  In the next section, we estimate the SNR expected to be measured by the LISA-Taiji network  and we study detection prospects of (chiral) GWs employing a Fisher-matrix forecast. 

To conclude this section we introduce the quantity 
\begin{equation}
	\label{eq:chi_def}
	\chi(f)\equiv \frac{P_R(f)-P_L(f)}{ I (f)} = \frac{V(f)}{I(f)}\,,
\end{equation}
which quantifies the strength of parity violation and is usually referred as chirality of the  SGWB (see, e.g.,~\cite{Alexander:2004wk}). By definition $|\chi| \leq 1$, and $|\chi|=0, 1$ correspond to zero/maximal parity violation respectively. Notice that we can rewrite $S_{ij}$ in terms of $\chi$ as
\begin{equation}
	\label{eq:two_point_dfreq3} 
	S_{ij}(f)   =   \frac{3 (H_0/h)^2}{4 \pi^2 f^3} \, \left[ h^2\Omega_{\rm GW}^I(f) \, \mathcal{R}^{I}_{ij}(f)  +h^2\Omega_{\rm GW}^V(f) \,  \mathcal{R}^{V}_{ij}(f)\right]  \; ,
\end{equation}
where we defined $\Omega_{\rm GW}^I(f) =  \Omega_{\rm GW}(f)$ and $ \Omega_{\rm GW}^V(f) =  \chi \, \Omega_{\rm GW}(f) $.

\section{The optimal Signal-To-Noise Ratio}
\label{sec:SNR}
Given the data streams $\tilde{d}_i$ for the two detectors (for the moment let us keep this general so that it applies both to self and cross-correlations), we can build a frequency integrated  estimator $\hat{\cal F}$ for the signal as~\cite{Seto:2008sr, Smith:2016jqs,Thorne:2017jft}
\begin{equation}
	\hat{\cal F}\equiv \sum_{i \geq j} \int \textrm{d}f_1\,\textrm{d}f_2  W^{ij}(f_1,\,f_2)\, \left[\tilde{d}_{i}(f_1)\,\tilde{d}_{j}(f_2)  -\frac{1}{2} \delta(f_1 + f_2) N_{ij}(f_1) \right] \,,
\end{equation}
where $N_{ij}$ is the noise power spectrum for channels $ij$ and $W^{ij}(f_1,\,f_2)$ is a filter function which satisfies $W^{ij}(f_1,\,f_2)^*=W^{ij}(-f_1,\,-f_2)$ and is $ij$ symmetric. Notice that the sum runs over $i \geq j$ to avoid double counting. The expectation value of $	\hat{\cal F}$ clearly reads
\begin{equation}
	\begin{aligned}
		\langle\hat{\cal F}\rangle &= \sum_{i \geq j}\int_{-\infty}^\infty \textrm{d}f_1\,\textrm{d}f_2\,W^{ij}(f_1,\,f_2) \, \left[\langle\tilde{d}_{i}(f_1)\,\tilde{d}_{j}(f_2)\rangle- \frac{1}{2} \delta(f_1 + f_2)  N_{ij}(f_1) \right] \\
		& =\frac{1}{2} \sum_{i \geq j}\,\int_{-\infty}^\infty \textrm{d}f_1 \textrm{d}f_2 \,W^{ij}(f_1,\,f_2)\,S_{ij}(f_1) \delta(f_1 + f_2)\,,
	\end{aligned}
\end{equation}
where $S_{ij}\left(f\right)$ is the signal power spectrum given in~\cref{eq:two_point_dfreq2}. 

In order to compute the SNR we have to evaluate the expectation value of $\hat{\cal F}$ over its variance in a noise-dominated regime, \emph{i.e.} 
\begin{equation}
	\label{eq:SNR_def}
\text{SNR} \equiv \frac{	\langle\hat{\cal F}\rangle} { \sqrt{ \left[ \langle \hat{\cal F}^2\rangle - \langle\hat{\cal F}\rangle^2 \right]_{S_{ij} = 0} } } \; .
\end{equation}
In order to evaluate the variance of our estimator, let us first compute the following four point correlation function~\footnote{The four-point correlators is expressed as a sum of products of two-point correlators using the Wick theorem} 
	\begin{equation}
		\begin{aligned}
			\tilde{\cal{F}}_{ijkl}^2 & \equiv	\left\langle \left( d_i(f_1) d_{j}(f_2) - \frac{1}{2}\delta(f_1 + f_2)  N_{ij}(f_1) \right)  \left( d_k(f_3) d_{l} (f_4)  - \frac{1}{2}\delta(f_3 + f_4)  N_{kl}(f_3) \right) \right\rangle\,, \\ 
			& =D_{ij}^{th}  D_{kl}^{th} + D_{ik}^{th}  D_{jl}^{th}  + D_{il}^{th}  D_{jk}^{th}   - \frac{1}{2}\delta(f_3 + f_4) D_{ij}^{th}  N_{kl}(f_3) + \\
			& \hspace{1cm}  - \frac{1}{2}\delta(f_1 + f_2) D_{lk}^{th}  N_{ij}(f_1) + \frac{1}{4}  N_{ij}(f_1) N_{lk}(f_3) \delta(f_1 + f_2) \delta(f_3 + f_4) \,,
		\end{aligned}
\end{equation}
where we used
	\begin{equation}
		\label{eq:D_Th}
		\langle d_i d_{j} \rangle = \langle s_i s_{j} \rangle + 	\langle n_i n_{j} \rangle \equiv 	D_{ij}^{th} = \frac{1}{2} \delta(f_1 +f_2 ) \left[  S_{ij} + N_{ij} \right]\; .
	\end{equation}
Since for $S_{ij} = 0 $ we clearly have $	\langle\hat{\cal F}\rangle = 0$ and $	\tilde{\cal{F}}_{ijkl}^2 $ reduces to
	\begin{equation}
	\begin{aligned}
\hspace{-1.3cm}	\left. 	\tilde{\cal{F}}_{ijkl}^2 \right|_{S_{ij} = 0 }& = \frac{1}{4}  N_{ik}(f_1) N_{jl}(f_2) \delta(f_1 + f_3) \delta(f_2 + f_4) +\frac{1}{4}  N_{il}(f_1) N_{jk}(f_3) \delta(f_1 + f_4) \delta(f_2 + f_3) \,,
	\end{aligned}
\end{equation}
the denominator of~\cref{eq:SNR_def} in the noise-dominated regime is directly given by
\begin{equation}
	\label{eq:F4_general}
	\begin{aligned}
	\left.	\langle \hat{\cal F}^2 \rangle \right|_{S_{ij} = 0 } =  & \sum_{i \geq j \, k \geq  l}\int \textrm{d}f_1\,\textrm{d}f_2\,\textrm{d}f_3\,\textrm{d}f_4\, W^{ij}(f_1,\,f_2)   W^{kl}(f_3,\,f_4) \, \left. \tilde{\cal{F}}_{ijkl}^2 \right|_{S_{ij} = 0 } \\
		=  & \frac{1}{4}\sum_{i \geq j \, k \geq  l} \,\int_{-\infty}^\infty \textrm{d}f_1 \textrm{d}f_2\,W^{ij}(f_1,\,f_2) W^{*kl}( f_1,\,f_2)\, \left[ N_{ik}(f_1)\,N_{jl}(f_2) +  N_{il}(f_1)\,N_{jk}(f_2) \right] \,. \\
	\end{aligned}
\end{equation}
We can now distinguish between self and cross-correlation. In the case of self-correlations $i$ and $j$ are both LISA (or Taiji) channels so that we have
\begin{equation}
	\label{eq:4point_self}
		\left.	\langle \hat{\cal F}^2 \rangle \right|_{S_{ij} = 0 } =   \frac{1}{2}\sum_{i \geq j } \,\int_{-\infty}^\infty \textrm{d}f_1 \textrm{d}f_2\,W^{ij}(f_1,\,f_2) W^{*ij}( f_1,\,f_2)\,  N_{ii}(f_1)\,N_{jj}(f_2)  \,, \\
\end{equation}
where we used the fact that $N_{ij}$ is diagonal in the AET/CDS bases and the two terms in the square bracket of~\cref{eq:F4_general} are equal. On the other hand for cross-correlation we have
\begin{equation}
	\label{eq:4point_cross}
	\left.	\langle \hat{\cal F}^2 \rangle \right|_{S_{mn} = 0 } =   \frac{1}{4}\sum_{m n} \,\int_{-\infty}^\infty \textrm{d}f_1 \textrm{d}f_2\,W^{m n}(f_1,\,f_2) W^{*m n}( f_1,\,f_2)\,  N_{m m}(f_1)\,N_{n n}(f_2)  \,, \\
\end{equation}
where the index $m$ runs over LISA channels and $n$ runs over Taiji channels. Notice that in this case one of the two terms in the square bracket of~\cref{eq:F4_general} is zero~\footnote{This assumes that all the components contributing to the noises of the two detectors are uncorrelated. While this is surely true for laser noises, this assumption must be tested for what concerns acceleration noise. }. Let us first focus on the case of self-correlations. The SNR in this case reads
\begin{equation}
	{\rm {SNR}_{self}}=   \frac{ \sum_{i \geq j}\,\int_{-\infty}^\infty \textrm{d}f_1 \textrm{d}f_2 \,W^{ij}(f_1,\,f_2)\,S_{ij}(f_1) \delta(f_1 +f_2)} {\left[2\sum_{i \geq j } \,\int_{-\infty}^\infty \textrm{d}f_1 \textrm{d}f_2\,W^{ij}(f_1,\,f_2) W^{*ij}( f_1,\,f_2)\,  N_{ii}(f_1)\,N_{jj}(f_2) \right]^{1/2}} \; .
\end{equation}
It's possible to show~\cite{Maggiore:1999vm} that, up to an irrelevant normalization constant, the SNR is maximized by 
\begin{equation}
	W^{ij} (f_1, f_2) \propto \,\delta(f_1 + f_2) \frac{S_{ij}(f_1)} {N_{ii}(f_1) N_{jj}(f_2)} \; .
\end{equation}
Finally the SNR for the self-correlation is given by~\footnote{The observation time appears after one of replaces one of the $\delta$s with a $\delta_T$ which keeps track of the finite observation time.}
\begin{eqnarray}
	\label{eq:snr_self}
	\hspace{-0.5cm}
	{\rm {SNR}_{self}}&=&\left[\sum_{i\geq j} \frac{T}{2} \int_{-\infty}^\infty \textrm{d}f_1\frac{S_{ij}(f_1)^2}{ \left[ N_{ii}(f_1)\,N_{jj}(f_2)\right]}\right]^{1/2}\nonumber\\
	&=& \left\{  T\, \sum_{i \geq j}\int_{0}^\infty df\;   \frac{\left[ h^2 \, \Omega^I_{\rm GW}(|f|) \; \mathcal{R}^{I}_{ij}(k)  +  \; h^2 \, \Omega^V_{\rm GW}(|f|) \; \mathcal{R}^{V}_{ij}(k) \right]^2} { N^{\Omega}_{ii}(|f|)  N^{\Omega}_{jj}(|f|) }\right\}^{1/2}\,,
\end{eqnarray}
where $T$ denotes the total observation time and we used eqs.~\eqref{eq:Omega_units} and~\eqref{eq:two_point_dfreq3} to express both the noise and the signal in $\Omega$ units. Since for self-correlations $\mathcal{R}^{V}_{ij}$ is zero, only $\Omega_{\rm GW}^I$ contributes to the SNR. 

On the other hand, with an analogous procedure, it is possible to show that for cross-correlations we get
\begin{equation}
	\label{eq:snr_cross}
\hspace{-0.5cm}
{\rm {SNR}_{cross}} =  \left\{ 2 T\, \sum_{m n}\int_{0}^\infty df\;   \frac{\left[ h^2 \, \Omega^I_{\rm GW}(|f|) \; \mathcal{R}^{I}_{m n}(k)  +  \; h^2 \, \Omega^V_{\rm GW}(|f|) \; \mathcal{R}^{V}_{m n}(k) \right]^2} { N^{\Omega}_{m m}(|f|)  N^{\Omega}_{n n}(|f|) }\right\}^{1/2}\,,
\end{equation}
where again the index $m$ runs over LISA channels and the index $n$ runs over Taiji channels. Notice the extra factor $2$ with respect to eq.~\eqref{eq:snr_self} which is due to the different prefactor in eq.~\eqref{eq:4point_cross} with respect to eq.~\eqref{eq:4point_self}. In this case since $\mathcal{R}^{V}_{m n}$ is non-zero, the SNR is sensitive to chiral contributions too.

We have seen that by using the self-correlations only $I$ contributes to the SNR.  By combining eqs.~\eqref{eq:snr_self} and~\eqref{eq:snr_cross} we can thus disentangle the contribution of $V$ in eq.~\eqref{eq:snr_cross}. Assuming a flat GW spectrum for $\Omega_V$ we can thus write
\begin{equation} \label{eq:SNRV_int}
	\hspace{-0.5cm}
	{\rm {SNR}_{V}} = h^2 \, \Omega^V_{\rm GW}(|f|)   \left\{ 2 T\,  \;  \sum_{i  j}\int_{0}^\infty df\;   \frac{   \left[ \mathcal{R}^{V}_{mn}(k) \right]^2} { N^{\Omega}_{m m}(|f|)  N^{\Omega}_{n n}(|f|) }\right\}^{1/2}\,,
\end{equation}
which becomes
\begin{equation} \label{eq:SNR_V}
	\hspace{-0.5cm}
	{\rm {SNR}_{V}}  \simeq  \chi \; \left( \frac{T}{3 \text{yrs}}\right)^{1/2} \,  \left( \frac{h^2 \, \Omega_{\rm GW}(|f|) }{2.50 \times 10^{-13} } \right)\; \,,
\end{equation}
where in performing the frequency integral of eq.~\eqref{eq:SNRV_int} we have used the same prescriptions for the noise and response functions as discussed in section~\ref{sec:network} and we used the definition of $\chi$ given in~\cref{eq:chi_def}. 
 
As discussed e.g. in~\cite{Smith:2016jqs}, the optimal configuration for the detection of chirality is obtained by taking the planes of the detectors (separated by a given distance $D$) to be parallel. In our case this would correspond to choosing $\theta_L+ \theta_T = \pi $ (\emph{i.e.} diametrically opposite) and parallel. By performing the integral in~\cref{eq:SNRV_int} for such case it's possible to show that the denominator in~\eqref{eq:SNR_V} would become $1.43 \times 10^{-13}$, which corresponds to an improvement of almost a factor 2 in the $\rm SNR$.

\section{Likelihood and Fisher forecasts}
\label{sec:fisher}
In this section we perform a Fisher forecast for both the intensity and the parity violating contributions. For this purpose we start by writing a likelihood function to describe the data in terms of some theoretical model. Using the definition in~\cref{eq:D_Th} we can easily compute the four point correlation function
	\begin{equation}
	\begin{aligned}
		\tilde{\cal{D}}_{ijkl}^2 & \equiv	\left\langle \left( d_i(f_1) d_{j}(f_2) - D_{ij}^{th} (f_1) \right)  \left( d_k(f_3) d_{l} (f_4) - D_{ij}^{th} (f_3) \right) \right\rangle \\ 
		& = \left\{ \frac{1}{4}  \left[ S_{ik}(f_1) + N_{ik}(f_1)  \right] \left[S_{jl}(f_2) + N_{jl}(f_2) \right] \delta(f_1 + f_3) \delta(f_2 + f_4) + \right. \\
	& \hspace{1cm}+  \left. \frac{1}{4}  \left[ S_{il}(f_1) + N_{il}(f_1)  \right] \left[S_{jk}(f_2) + N_{jk}(f_2) \right] \delta(f_1 + f_4) \delta(f_2 + f_3) 	\right\} \; .
	\end{aligned}
\end{equation}
Let us directly work in $\Omega$ units so that, similarly to~\cite{Caprini:2019pxz, Flauger:2020qyi}, we can build our likelihood as
\begin{equation}
	\begin{aligned}
		- \ln \mathcal{L}  & = \frac{N_c}{2} \sum_{\Lambda \Lambda'} \sum_k w^{k}_{\Lambda \Lambda'} \frac{\left[ D_{\Lambda \Lambda'}^{k} - D^{th}_{\Lambda \Lambda'} (f_k) \right]^2} {\sigma_{s,\Lambda \Lambda'}^2}    +  \frac{N^{'}_c}{2}  \sum_{\tau \tau'} \sum_k w^{k}_{\tau \tau'} \frac{ \left[ D_{\tau \tau'}^{k} - D^{th}_{\tau \tau'} (f_k) \right]^2} {\sigma_{s,\tau \tau'}^2}    +\\ 
		&   +  \frac{N^{''}_c}{2}  \sum_{\Lambda \tau} \sum_k w^{k}_{\Lambda \tau} \frac{\left[  D_{\Lambda \tau}^{k} - D^{th}_{\Lambda \tau} (f_k)\right]^2 } {\sigma_{c,\Lambda \tau}^2}\,,
	\end{aligned}
\end{equation}
where $N_c,N^{'}_c,N^{''}_c$ denote the number of data segments (chunks) used in the analysis\footnote{This is directly set by the choice of the segmentation of the time domain data stream to convert it  into frequency domain. In principle these are different for the two experiments and in particular $N_c^{''}$, number of data segments we can use for the correlation, could be different from either $N_c$ and $N_c^{'}$. For the sake of simplicity in the following we will ultimately assume  $N_c = N^{'}_c = N^{''}_c$. }, the apex $k$ runs over frequencies, $ D^{k}_{ij}$ here denotes the average over data segments of $ 3 (H_0/h)^2 d^k_i d^k_j  / (4 \pi^2 f_k^3) $, we denote with $w^k_{ij}$ the weights associated to the point (in the $ij$ channels combination) at frequency $f_k$, the indexes $\Lambda,\Lambda'$ run over noise-diagonal LISA channels, the indexes $\tau,\tau'$ run over noise-diagonal Taiji channels and we introduced
\begin{equation}
	\begin{aligned}
		\sigma_{s,\Lambda\Lambda'}^2 & =\left( \sum_{\lambda} \mathcal{R}^\lambda_{\Lambda\Lambda} \, h^2 \, \Omega^\lambda_{\rm GW} + N^\Omega_{\Lambda\Lambda} \right) \left( \sum_{\lambda} \mathcal{R}^\lambda_{\Lambda'\Lambda'} \, h^2 \, \Omega^\lambda_{\rm GW} + N^\Omega_{\Lambda'\Lambda'}\right) \; , \\ 
		\sigma_{s,\tau \tau }^2 &  =\left( \sum_{\lambda} \mathcal{R}^\lambda_{\tau \tau } \, h^2 \, \Omega^\lambda_{\rm GW} + N^\Omega_{\tau \tau }\right) \left( \sum_{\lambda} \mathcal{R}^\lambda_{\tau' \tau' } \, h^2 \, \Omega^\lambda_{\rm GW} + N^\Omega_{\tau'\tau' }\right) \; ,  \\	
		\sigma_{c,\Lambda\tau}^2 & =\frac{1}{2} \left[  \sum_{\lambda, \lambda' }\mathcal{R}^\lambda_{\Lambda\tau} \mathcal{R}^{\lambda'}_{\Lambda\tau} \, h^2 \, \Omega^\lambda_{\rm GW} \, h^2 \, \Omega^{\lambda{\rm GW}}_h  +   \right. \\
	&  \hspace{1cm}	+ \left. \left( \sum_{\lambda} \mathcal{R}^\lambda_{\Lambda\Lambda} \, h^2 \, \Omega^\lambda_{\rm GW}  + N^\Omega_{\Lambda\Lambda}\right) \left( \sum_{\lambda} \mathcal{R}^\lambda_{\tau\tau} \, h^2 \, \Omega^\lambda_{\rm GW} + N^\Omega_{\tau\tau}\right) \right] \; .
	\end{aligned}
\end{equation}
Notice that similarly to~\cref{eq:4point_self} we summed over the two identical contributions in the self-correlations. Using the definitions of the Stokes parameters and of the corresponding responses given in~\cref{eq:stokes} and~\cref{eq:resp_stokes}, we can directly use $I$ and $V$ in $\Omega$ units as given in~\cref{eq:two_point_dfreq3} as parameters to model the data. In particular, since $\mathcal{R}^V$ is zero for the self-correlations, it's easy to show that
\begin{eqnarray}
	D^{th}_{\Lambda \Lambda'} &=& h^2 \, \Omega^I_{\rm GW}  \mathcal{R}^I_{\Lambda \Lambda'} + N^\Omega_{\Lambda \Lambda'} \; ,  \\D^{th}_{\tau \tau'} &=& h^2 \, \Omega^I_{\rm GW}  \mathcal{R}^I_{\tau \tau'} + N^\Omega_{\tau \tau'} \; , \\
	  D^{th}_{\Lambda \tau} &=& h^2 \, \Omega^I_{\rm GW}  \mathcal{R}^I_{\Lambda \tau'} + h^2 \, \Omega^V_{\rm GW}  \mathcal{R}^V_{\Lambda \tau'}  \; .
\end{eqnarray}
This clearly shows that self-correlations only contribute to the measurement of $I$ while, as expected, cross-correlations also contribute to $V$. Let us proceed by directly using $ \Omega^I_{GW}$ and $\Omega^V_{GW}$ as free parameters to model the data \emph{\emph{i.e.}} defining $\theta$ the vector of parameters we have $\theta_\alpha = \{h^2 \, \Omega^I_{GW}, h^2 \, \Omega^V_{GW}\}$. As customary the best fit $\bar{\theta}$ is defined by maximizing the likelihood as
\begin{equation}
	- \left. \frac{\partial \ln \mathcal{L} }{\partial \theta_\alpha }   \right|_{\theta = \bar{\theta}} =  0  \; , 
\end{equation}
and the Fisher matrix as
\begin{eqnarray}
	\label{eq:fisher}
	F_{\alpha \beta} &\equiv& -  \left. \left\langle \frac{ \partial^2 \ln \mathcal{L} }{\partial \theta_\alpha \partial \theta_\beta } \right\rangle \right|_{\theta = \bar{\theta}} \\
	& =&  \delta_{\alpha I} \delta_{\beta I} \left[ N_c \sum_{\Lambda \Lambda'}  \sum_k  \frac{w^{k}_{\Lambda \Lambda'} } {\sigma_{s,\Lambda \Lambda'}^2} \left[ \mathcal{R}^I_{\Lambda \Lambda'}  \right]^2  + N^{'}_c \sum_{\tau \tau'}  \sum_k  \frac{w^{k}_{\tau \tau'} } {\sigma_{s,\tau \tau'}^2}  \left[ \mathcal{R}^I_{\tau \tau'}  \right]^2 \right] \\ 
		&& \hspace{0.5cm} + N^{''}_c   \sum_{\Lambda \tau} \sum_k  \frac{w^{k}_{\Lambda \tau} } {\sigma_{s,\Lambda \tau}^2}  \frac{\partial D^{th}_{\Lambda \tau} (f_k)}{\partial \theta_\alpha }    \frac{\partial D^{th}_{\Lambda \tau} (f_k)}{\partial \theta_\beta }  \Bigg|_{\theta = \bar{\theta}} \; .
	\end{eqnarray}
Once again, this manifestly shows that self-correlations only contribute to the $II$ entry and the cross-correlations also contributes to the $IV$ and $VV$ entries as well. 

\begin{figure}[t!]
	\hspace{-1cm}
	\includegraphics[width=1.1\linewidth]{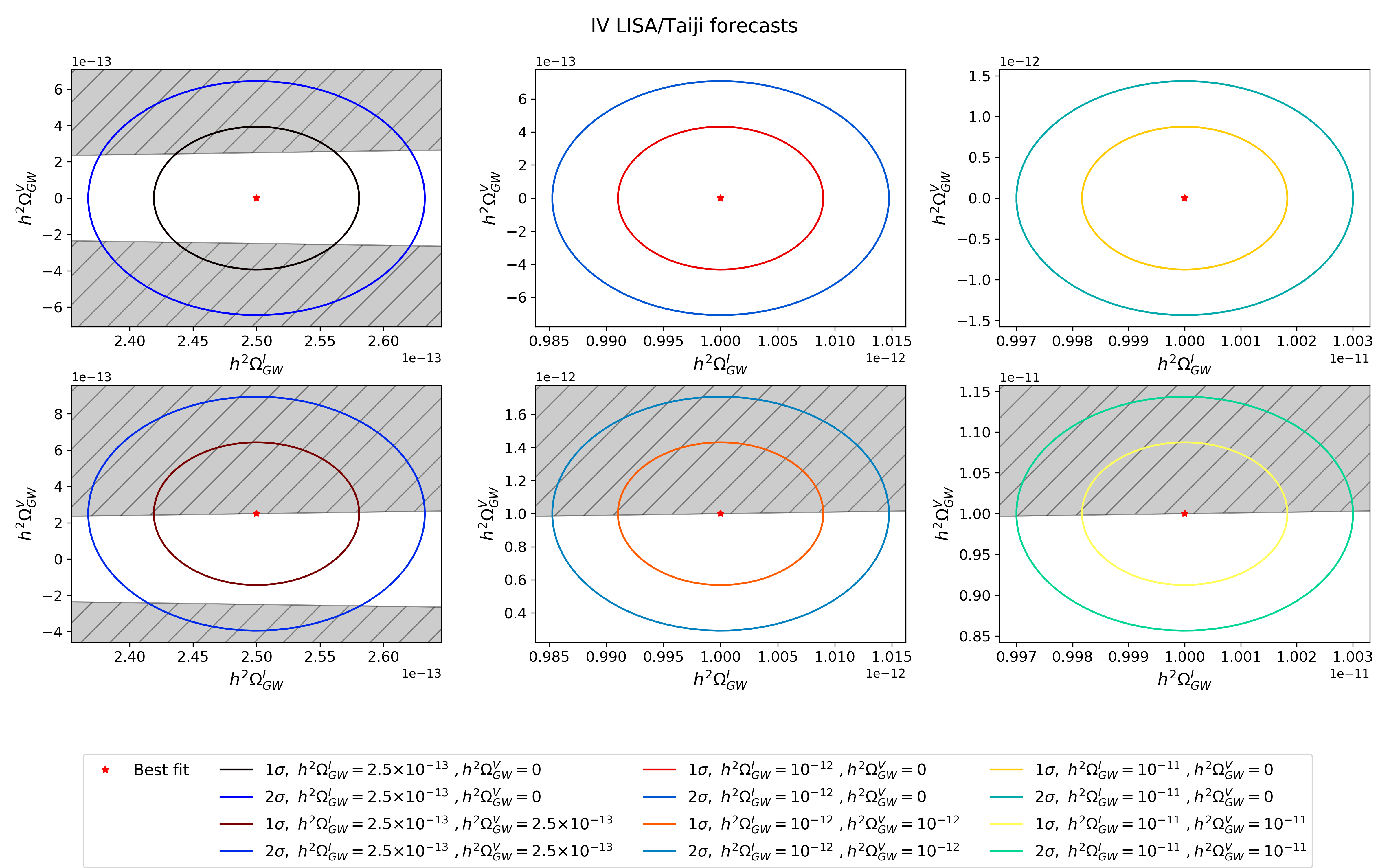} 
	\caption{ \it Joint $IV$ forecasts using the LISA/Taiji network for six different choices of $h^2 \, \Omega^I_{GW}$, $h^2 \, \Omega^V_{GW}$. The three columns respectively correspond to $h^2 \, \Omega^I_{GW} = [2.5 \times 10^{-13},10^{-12}, 10^{-11} ] $ and the two rows are $h^2 \, \Omega^V_{GW}=[ 0, h^2 \, \Omega^I_{GW} ] $. The gray shaded areas correspond to regions of the parameter space with $h^2 \, \Omega^I_{GW}< |h^2 \, \Omega^V_{GW}|$ which is theoretically unacceptable. \label{fig:forecasts} }
\end{figure}

To conclude this section we present in~\cref{fig:forecasts} the results obtained by assuming that LISA and Taiji will perform their measurement simultaneously and with the same data segmentation (implying $N_c = N^{'}_c = N^{''}_c$) for a total observation time of $4$ years with $75\%$ efficiency. Assuming time segments of roughly $11.5$ days (leading to a maximal frequency resolution of $\sim 10^{-6}$Hz) we obtain $\sim 95$ data segments during the whole missions' duration. The plots show the 1 and $2\sigma$ regions (respectively in blue and orange) for the best-fit parameters (represented by a red star) in the $h^2 \, \Omega^I_{\rm GW}  - h^2 \, \Omega^V_{\rm GW}$ planes. The gray shaded areas, while being perfectly viable in the  Bayesian framework depicted by our likelihood, are theoretically inconsistent since they correspond to $|\chi|>1$, where $\chi $ is the quantity introduced in~\cref{eq:chi_def}. By looking at the second row, where the signal is chosen to be maximally chiral, \emph{\emph{i.e.}} $h^2 \, \Omega^I_{\rm GW} = h^2   \Omega^V_{\rm GW}$ , it possible to notice that the estimated error on $h^2 \, \Omega^V_{\rm GW}$ is roughly between 1 and 2 orders of magnitude larger than the estimated error on $h^2 \, \Omega^I_{\rm GW}$. This can be explained by noticing that, as shown in~\cref{eq:fisher}, the $F_{\alpha \beta}$ is proportional to the response squared. Since the plots of $\mathcal{R}^I$ and $\mathcal{R}^V$ shown in~\cref{fig:response} show that the maximum of $\mathcal{R}^I$ is roughly 1 order of magnitude larger than the maximum of $\mathcal{R}^V$ (which however is located at larger frequencies), the error bars in~\cref{fig:forecasts} are order-of-magnitude consistent. As a last comment, the bottom left plot show that, for a maximally chiral signal with $h^2 \, \Omega^I_{\rm GW} = 2.5 \times 10^{-13}$ it will not be possible to claim a detection of parity violation. While, on the other hand, a detection is clearly possible for maximally chiral signals with $h^2 \, \Omega^I_{GW}= 10^{-12}$ (bottom central plot). Therefore, the minimal value for detecting $h^2 \, \Omega^V_{\rm GW}$ for maximally chiral signals should sit somewhere between these two values for $h^2 \, \Omega^I_{\rm GW}$.

\begin{figure}[t!]
	\hspace{-1cm}
	\includegraphics[width=1.1\linewidth]{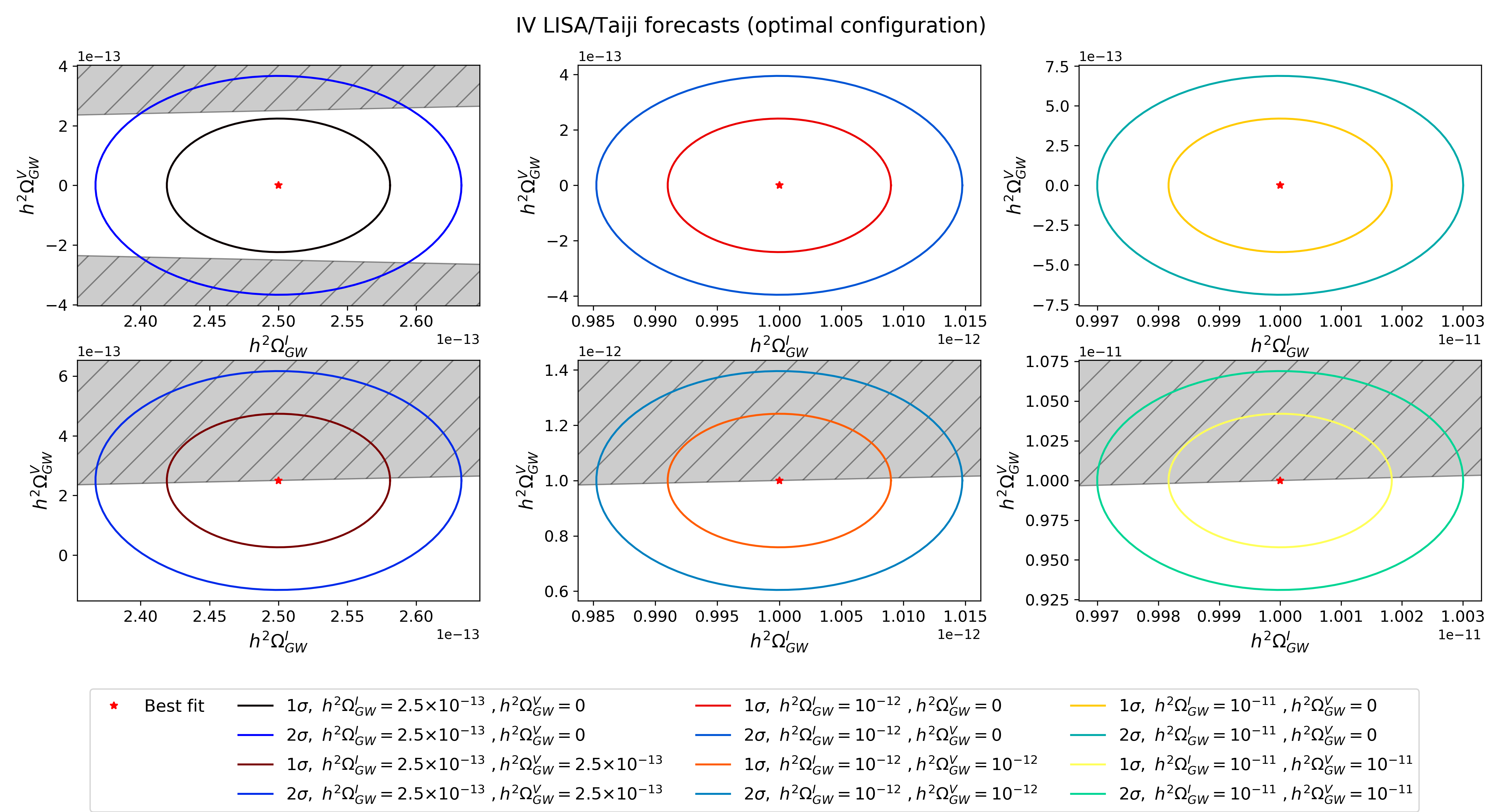} 
	\caption{ \it Same plot structure and injected signals as~\cref{fig:forecasts} but with Taiji at its optimal location for detecting chirality (see text for details). \label{fig:forecastsopt} }
\end{figure}

Finally, in~\cref{fig:forecastsopt} we show the same forecasts as in~\cref{fig:forecasts} but placing Taiji at its optimal location for detecting chirality \emph{i.e.} $\theta_L+ \theta_T = \pi $. We clearly notice an improvement with respect to~\cref{fig:forecasts} and in particular, by comparing the bottom bottom left plots of these two figures, we see that in the optimal case we have a $1\sigma$ detection of chirality and we start to disfavor some values for $h^2 \, \Omega_{\rm GW}^V$ at $2 \sigma$ . In general, by comparing the error bars for $h^2 \, \Omega^V_{\rm GW}$ in~\cref{fig:forecasts} and~\cref{fig:forecastsopt}, we see approximately a factor two improvement in the determination of this quantity. This is consistent with the estimate for the SNR performed in~\cref{sec:SNR}.

\section{Conclusions}
\label{sec:conclusions}
In this paper we have estimated the sensitivity of the LISA-Taiji detector network to a chiral isotropic SGWB background, taking into account self and cross-correlation of all the channels of the two interferometers. LISA and Taiji are two space-based detectors which will be sensitive in the milli-Hertz regime and will probably fly at the same time, around 2034. While independently the two detectors will be unable to detect parity violation of an isotropic SGWB, due to their planar configuration, in this paper we have shown that, cross-correlating the output of the two detectors we will be able constrain the Stokes parameter $V$, which characterizes the asymmetry between the amplitudes of the left- and right-handed polarized waves. 

We have computed the network response functions to a SGWB working in the TDI 1.5 channel basis, considering the expected Taiji orbit specifications and using the most up-to-date LISA instrument specifications. We have then estimated the significance of such a detection through a SNR estimator, finding that a clear measurement of chirality can be claimed for a maximally chiral signal with $h^2 \, \Omega^I_{\rm GW} \simeq 10^{-12}$, in accordance with a recent similar analysis.

 Finally, as a novel contribution, we have also performed performed a Fisher analysis to estimate the errors on the $I$ and $V$ parameters in the case of an un-polarized background and in the case of a maximally chiral background.  In the second case, we have seen that the estimated error on $V$ is roughly between 1 and 2 orders of magnitude larger than the estimated error on $I$. We have motivated this from the fact that the Fisher matrix $F_{\alpha \beta}$ is proportional to the response squared and, since the maximum of $\mathcal{R}^I$ is roughly 1 order of magnitude larger than the maximum of $\mathcal{R}^V$, this can explain the error bars order-of-magnitude difference. This can be graphically seen from the response plots $\mathcal{R}^I$ and $\mathcal{R}^V$ shown in~\cref{fig:response}.

Even if our analysis has been performed under the assumption of a flat spectrum ($n_T = 0$) for stochastic gravitational waves, we argue that our final claims are general. Models of inflation predicting detectable GW at the frequency scales of space-based interferometers are accompanied with a blue-tilted GW spectrum, $n_T>0$ (see, e.g., \cite{Bartolo:2016ami}). However, the cross-correlations considered here are sensitive to GW in a relatively small frequency window where a soft scale dependence can be neglected if one is interested to have a model independent order-of-magnitude estimate. Conversely, for strong scale dependence one can adapt our analysis to specific models of inflation in order to obtain model dependent exact results.

Our analysis can be extended to include also other detector satellites, like for instance TianQin~\cite{Luo:2015ght}, which is another Chinese space-based detector planned to be launched in the second half of the next decade. We leave this for future works.

\acknowledgments

We thank Jesus Torrado for a careful reading of the draft and for insightful comments.
The work of M.P. was supported by STFC grants ST/P000762/1 and ST/T000791/1. G.O. acknowledges support from the Netherlands organization for scientific research (NWO) VIDI grant (dossier 639.042.730).
M.P. acknowledges support by the European Union’s Horizon 2020 Research Council grant 724659 MassiveCosmo ERC–2016–COG. A.R. acknowledges funding from Italian Ministry of Education, University and Research (MIUR) through the `Dipartimenti di eccellenza' project Science of the Universe.

\appendix
\section{Useful formulae}
\label{sec:appendix}
In this appendix we briefly summarize the key ingredients for the definition of the TDI variables and for the computation of the detector response function. We closely follow the notation of Appendix A of~\cite{Flauger:2020qyi} to which we remand for further details. 
\subsection{GW expansion}
\label{sec:formulae}
Let us work in natural units and in the Lorentz transverse-traceless gauge. Given a coordinate system $\{\hat e_x, \hat e_y, \hat e_z\}$ arbitrarily oriented and at rest with respect to the isotropic SGWB we are considering, for a single wave-vector $\mathbf{k} $ of an incoming plane GW  we can define the following orthonormal basis~\cite{Bartolo:2018qqn,Domcke:2019zls}
\begin{equation}
	\hat{u}(\hat{k}) = \frac{\hat{k} \times \hat{e}_z}{|\hat{k} \times \hat{e}_z|} \; , \hspace{4cm} \hat{v} (\hat{k} )= \hat{k} \times \hat{u} \;, 
\end{equation}
where $\hat k$ is unit vector in the direction of $\mathbf{k}$ and we will denote its magnitude by $k=|\mathbf{k}|$. Using this basis, we can define the so-called ``plus" ($+$) and ``cross" ($\times$) polarization tensors
\begin{equation}
	\label{eq:PlusCrossTensors}
	e_{ab}^{(+)} (\hat{k})= \frac{{\hat u}_a {\hat u}_b - {\hat v}_a {\hat v}_b }{\sqrt{2}}  \; , \hspace{4cm} 
	e_{ab}^{(\times)}(\hat{k}) = \frac{{\hat u}_a {\hat v}_b + {\hat v}_a {\hat u}_b }{\sqrt{2}} \;. 
\end{equation} 
Since $\hat{u} (- \hat{k} ) = - \hat{u} (\hat{k} )$ and  
$\hat{v} (- \hat{k} ) = \hat{v} (\hat{k})$, 
the tensors $e_{ab}^{(+)}$ and $
e_{ab}^{(\times)}$ fulfill the following conditions
\begin{equation}
	\begin{aligned}
		& e^{+/\times}_{ab} (\hat{k} )  = e^{+/\times \, *}_{ab} (\hat{k})  \;,  \hspace{1cm} 
		&& e^{+}_{ab} (\hat{k} )   = e^{+}_{ab} (- \hat{k} )   \;,  \hspace{1cm} 
		&& e^{\times}_{ab} (- \hat{k} )  = - e^{\times}_{ab} (\hat{k} ) \; ,\\
		& e^+_{ab} (\hat{k})  e^+_{ab}(\hat{k}) = 1 \; , 
		&& e^\times_{ab} (\hat{k}) e^\times_{ab}(\hat{k})  = 1 \;, &&  e^+_{ab} (\hat{k}) e^\times_{ab} (\hat{k})  = 0 \; .
	\end{aligned}
\end{equation}
Analogously, we introduce the ``right-handed" (R) and ``left-handed" (L) polarization tensors
\begin{equation}
	e_{ab}^{R} (\hat{k} ) = 
	\frac{ {\hat u}_a  + i \,  {\hat v}_a  }{\sqrt{2}} \, 
	\frac{ {\hat u}_b  + i \,  {\hat v}_b  }{\sqrt{2}} 
	\;, \hspace{0.5cm} 
	e_{ab}^{L} (\hat{k} ) = 
	\frac{ {\hat u}_a  - i \,  {\hat v}_a  }{\sqrt{2}} \, 
	\frac{ {\hat u}_b  - i \,  {\hat v}_b  }{\sqrt{2}} 
	\ \;.  
	\label{u-v-oneindex} 
\end{equation} 
These are related to the plus and cross polarization basis by means of the relationships
\begin{eqnarray}\label{eq:ChiralTensors}
	e_{ab}^{R} (\hat{k} ) = \frac{e_{ab}^{+} + i \, e_{ab}^{\times}}{\sqrt{2}}  \;, \hspace{1cm} 
	e_{ab}^{L} (\hat{k} ) = \frac{e_{ab}^{+} - i \, e_{ab}^{\times}}{\sqrt{2}}  \;.
\end{eqnarray} 
We can express the superposition of all GWs reaching the position $x$ at the time $t$ in terms of incoming plane waves as
\begin{equation}
	\label{eq:h_of_t_k}
	h_{ab}(\mathbf{x},t)  = \int_{- \infty}^{+ \infty} \textrm{d} f \int_{\Omega} \textrm{d} \Omega_{\hat{k} }\; \textrm{e}^{2\pi i f (t - \hat{k} \cdot \mathbf{x})}  \; \sum_{P} \tilde{h}_{P} (f, \hat{k})  \; e_{ab}^{P}(\hat{k}) \;  ,
\end{equation}
where $P$ is the GW polarization (either $+/\times$ or $L/R$), $f = k/2 \pi$ is the frequency of each plane wave, $\textrm{d} \Omega_{\hat{k}}$ is the infinitesimal solid angle from which the incoming wave with wave-vector $\mathbf{k}$ arrives, and finally $\tilde{h}_{P} (f, \hat{k}) \equiv  f^2 \, \tilde{h}_{P} (\mathbf{k}) $. In terms of $ \mathbf{k} $ we can thus write
\begin{equation}
	h_{ab}(\mathbf{x},t)  =   \int \textrm{d}^3 k  \; \textrm{e}^{- 2\pi i \mathbf{k}\cdot \mathbf{x} } \sum_{P}  \left[  \textrm{e}^{2\pi i k t }  \; \tilde{h}_{P} (\mathbf{k})  \; e_{ab}^{P}(\hat{k})  +  \textrm{e}^{-2\pi i k t }  \; \tilde{h}_{P}^* (-\mathbf{k})  \; e_{ab}^{P \, *}(-\hat{k})\right] \; .
\end{equation}
From now on we will use the $L/R$ basis which will be denoted with $\lambda$.

\subsection{TDI variables and response functions} \label{app:response}
In the following, we neglect the motion of the detector satellite (either LISA or Taiji). Consider two test masses labelled 1 and 2, inside two of the three detector spacecrafts, located at $\mathbf{ x}_1$ and $\mathbf{x}_2$ and separated by the vector $L \hat{l}_{12}$, with $ \hat{l}_{12}= (\mathbf{x}_2 - \mathbf{x}_1 )/ |\mathbf{ x}_1 - \mathbf{x}_2 |$. In the following we assume $|\mathbf{ x}_1 - \mathbf{x}_2 |$ to be constant (and  equal to $L=2.5\times 10^9\,$m for LISA or $L=3\times 10^9\,$m for Taiji) . When a GW crosses the detector, a photon leaving test mass $2$ at time $t-L$ is received at test mass $1$ with a time shift~\cite{estabrook, Romano:2016dpx}
\begin{equation}
	\Delta T_{12}(t) = \frac{\hat{l}^a_{12} \hat{l}^b_{12} }{2 } \int_{0}^{L} \textrm{d}s \, h_{ab} (t(s), \mathbf{x}(s) ) \; .
\end{equation}
At the lowest order the photon path satisfies $t(s)=t-L+s$ and $\mathbf{ x}(s)= \mathbf{ x}_2 -s \hat{l}_{12} $. It follows
\begin{equation}
	\label{eq:time_shift}
	\begin{aligned}
		\Delta T_{12}(t) =  L \int \textrm{d}^3 k \; \textrm{e}^{ -2\pi i \mathbf{k} \cdot \mathbf{x}_2 }  \sum_{\lambda=L,R} 
		&
		\left[ \textrm{e}^{2\pi i k (t-L) } 	\mathcal{M} (\mathbf{k}, \hat{l}_{12})  \; \tilde{h}_{\lambda } (\mathbf{k})  \; \mathcal{G}^{\lambda }(\hat{k}, \hat{l}_{12})  \right. \\
		&
		+ \left. \textrm{e}^{-2\pi i k (t-L) } 	\mathcal{M}^* (-\mathbf{k}, \hat{l}_{12})  \;  \tilde{h}^*_{\lambda} (-\mathbf{k})  \; \mathcal{G}^{\lambda \ *}(-\hat{k}, \hat{l}_{12}) \right]  \; ,
	\end{aligned}
\end{equation}
where we introduced the two quantities
\begin{equation}
	\mathcal{G}_i^A(\hat{k}, \hat{l}_{ij})  \equiv  \frac{\hat{l}^a_{ij} \hat{l}^b_{ij} }{2} \, e_{ab}^{A}(\hat{k})  \; , \qquad 
	\mathcal{M} (\mathbf{k}, \hat{l}_{ij}) \equiv \textrm{e}^{\pi i k L (1 + \hat{k} \cdot \hat{l}_{ij} ) }  \text{sinc}\left[ \pi k L  (1 + \hat{k} \cdot \hat{l}_{ij} ) \right] \; ,
\end{equation}
and the definition of sinc: $\text{sinc} (x) \equiv  \sin(\pi x) / (\pi x).$ \\

In practice LISA will not measure time shifts but rather differential Doppler frequency shifts defined as $\Delta F_{12} (t) \equiv \Delta \nu_{12} (t)/\nu = -\textrm{d} \Delta T_{12} (t)/\textrm{d} t $. Moreover, given the uncertainty on the mass positions and on the laser frequency, LISA employs Time Domain Interferometry (TDI) techniques. This requires the usage of the same laser light pulse split into different path to enforce better noise control. Since the techniques to be adopted are still under discussion, in the following we chose simplified light paths which contain the keys ingredients which will ultimately be used. Without loss of generality, in the following we assume Taiji will use the same observable (\emph{\emph{i.e.}} differential Doppler frequency shifts) and similar TDI techniques as the ones to be adopted in LISA.\\

The simplest interferometric measurement one could define is just
\begin{equation}
	\Delta F_{1(23)}(t) \equiv  \Delta F_{21}(t -L) +  \Delta F_{12}(t)  - \left[  \Delta F_{31}(t -L) +  \Delta F_{13}(t)  \right] \; .
	\label{eq:tdi1}
\end{equation}
By substituting~\cref{eq:time_shift}
into~\cref{eq:tdi1}, one can see that the signal contribution to $\Delta F_{1(23)}(t)$ is 
\begin{equation}
	\begin{aligned}
		\Delta F_{1(23)}(t)	&  = - \int \textrm{d}^3 k \; \textrm{e}^{ -2\pi i \mathbf{k} \cdot \mathbf{x}_1 } (2\pi k L) \sum_{\lambda} \left[  \textrm{e}^{2\pi i k (t -L) } R_1^{\lambda}(\mathbf{k}, \hat{l}_{12}, \hat{l}_{13} )  \tilde{h}_{\lambda} (\mathbf{k})  +  \right. \\ 
		& \hspace{4cm}- \left. \textrm{e}^{-2\pi i k (t -L) } {R_1^{\lambda}}^*(-\mathbf{k}, \hat{l}_{12}, \hat{l}_{13} )  \tilde{h}_{\lambda}^* (-\mathbf{k})\right] \; , 
	\end{aligned} 
\end{equation}
where we introduced the functions $R_i^{\lambda}$
\begin{equation}
	\label{eq:R_func}
	R_i^{\lambda}(\mathbf{k}, \hat{l}_{ij}, \hat{l}_{ik} ) \equiv   \mathcal{G}^{\lambda }(\hat{k}, \hat{l}_{ij})  \mathcal{T}(\mathbf{k}, \hat{l}_{ij}) - \mathcal{G}^{\lambda }(\hat{k}, \hat{l}_{ik})  \mathcal{T}(\mathbf{k}, \hat{l}_{ik})  \; ,
\end{equation}
and the so-called detector transfer function $\mathcal{T}$ defined as
\begin{equation}
	\label{eq:transfer_function}
	\hspace{-0.5cm}
	\mathcal{T} (\mathbf{k}, \hat{l}_{ij})  \equiv \textrm{e}^{\pi i k L (1 - \hat{k} \cdot \hat{l}_{ij} ) }  \text{sinc}\left[\pi k L  (1 + \hat{k} \cdot \hat{l}_{ij} ) \right] + \textrm{e}^{- \pi i k L (1 + \hat{k} \cdot \hat{l}_{ij} ) }  \text{sinc}\left[ \pi k L  (1 - \hat{k} \cdot \hat{l}_{ij} ) \right] \; .
\end{equation}
Introducing the detector characteristic frequency $f_* \equiv (2 \pi L)^{-1}$, we can interpret the detector transfer function as a low-pass filter, which is nearly constant for $f \ll f_*$ and rapidly decays as the GW frequency becomes larger than $f_*$. The functions $R_i^{\lambda}$ also contain geometrical information and describe how the detector responds to a plane wave of wave-vector $\mathbf{ k}$ when the detector test masses $i$ and $j$ are oriented along the direction $\hat{l}_{ij}$. A more robust TDI variable is the so-called TDI 1.5 X variable (TDI variables $\text{Y}$ and $\text{Z}$ are obtained as well by cyclic permutation), defined as~\cite{Flauger:2020qyi}
\begin{equation}
	\hspace{-1.2cm}
	\begin{aligned}
		\Delta F^{1.5}_{1(23)}(t) & = \Delta F_{1(23)}(t -2L) + \Delta F_{1(32)}(t)   \\
		\label{eq:tdi15}
		&  = - \int \textrm{d}^3 k \; \textrm{e}^{ -2\pi i \textbf{k} \cdot \mathbf{x}_1 } (2\pi i k L) \sum_{\lambda} \left[  \textrm{e}^{2\pi i k (t -L) } W(k L) R_1^{\lambda}(\mathbf{k}, \hat{l}_{12}, \hat{l}_{13} )  \tilde{h}_{\lambda} (\mathbf{k})  +  \right. \\ 
		& \hspace{4.8cm}- \left. \textrm{e}^{-2\pi i k (t -L) }W^*(k L) {R_1^{\lambda}}^*(-\mathbf{k}, \hat{l}_{12}, \hat{l}_{13} )  \tilde{h}_{\lambda}^* (-\mathbf{k})\right] \; , 
	\end{aligned} 
\end{equation}
with $W(k L) \equiv  \textrm{e}^{- 4\pi i k  L } -1 $. In order to ease the notation, from now on we use the simplified notation $s_i(t) \equiv \Delta F^{1.5}_{i(jk)}(t)$. Notice that we also drop the two indexes $jk$ since in each detector they could trivially be determined (via cyclic permutations) given $i$.

As explained in~\cref{sec:network}, the information is contained in two-point correlation functions of data streams. In the following, we proceed in complete generality, \emph{i.e.} without assuming that the two data streams are from the same detector. Let us denote with $L_i$ ($L_l$) the armlength of the detector associated with $s_i(t)$ ($s_j(t)$) so that:
\begin{equation}
	\label{eq:two_point_dfreq}
	\begin{aligned}
	\hspace{-0.5cm}	\left\langle s_i(t) s_j(t) \right\rangle & =   \int \textrm{d}^3  k \; \textrm{e}^{ -2\pi i  \mathbf{k} \cdot (\mathbf{x}_i - \mathbf{x}_j) }  (2\pi k L_i) (2\pi k L_j) \, \sum_{\lambda} \frac{ P_{\lambda} (k) }{4 \pi k^2}   \times \\
		& \hspace{1cm} \times \left[ \textrm{e}^{-2 \pi i k (L_i -L_j)} W(k L_i) \, W^*(k L_j)    R_{i}^{\lambda}(\mathbf{k}, \hat{l}_{ik}, \hat{l}_{il} ) {R_{j}^{\lambda}}^*(\mathbf{k}, \hat{l}_{jm}, \hat{l}_{jn} )   +  \right. \\
		& \hspace{1cm} \times \left. \textrm{e}^{2 \pi i k (L_i -L_j)} W^*(k L_i) \, W(k L_j)    {R_{i}^{\lambda}}^* (-\mathbf{k}, \hat{l}_{ik}, \hat{l}_{il} ) R_{j}^{\lambda}(-\mathbf{k}, \hat{l}_{jm}, \hat{l}_{jn} )  \right] \; ,
	\end{aligned}
\end{equation}
where we have used the definition of GW power spectrum given in~\cref{eq:spectrum}. Notice that for $L_i = L_j = L$ we get $|W(kL)|^2 = 2 \left[ 1 - \cos(4\pi k L) \right] = 4 \sin^2(2\pi k L)$. By using the symmetry properties of  $\mathcal{T}$ it is possible to show that ${R_{i}^{\lambda}}^* (-\mathbf{k}, \hat{l}_{ik}, \hat{l}_{il} ) R_{j}^{\lambda}(-\mathbf{k}, \hat{l}_{jm}, \hat{l}_{jn} ) = R_{i}^{\lambda}(\mathbf{k}, \hat{l}_{ik}, \hat{l}_{il} ) {R_{j}^{\lambda}}^*(\mathbf{k}, \hat{l}_{jm}, \hat{l}_{jn} )$, so that~\eqref{eq:two_point_dfreq} reduces to:
\begin{equation}
	\begin{aligned}
		\hspace{-0.5cm}	\left\langle s_i(t) s_j(t) \right\rangle & =   \int \textrm{d}  k \; (2\pi k L_i) (2\pi k L_j) \, \sum_{\lambda}  P_{\lambda} (k)    \left[ \textrm{e}^{-2 \pi i k (L_i -L_j)} W(k L_i) \, W^*(k L_j)    	\tilde{R}^{\lambda}_{ij, \,(kl)(mn)}(k) + h.c.\right] \; ,
	\end{aligned}
\end{equation}
where we defined (again we drop the indexes between parenthesis)
\begin{equation}
	\tilde{R}^{\lambda}_{ij} (k) \equiv \frac{1}{4 \pi }\int \textrm{d}^2 \hat{k} \;  \textrm{e}^{ -2\pi i   \mathbf{k} \cdot (\mathbf{x}_i - \mathbf{x}_l) }  R_{i}^{\lambda}(\mathbf{k}, \hat{l}_{ik}, \hat{l}_{il} ) {R_{j}^{\lambda}}^*(\mathbf{k}, \hat{l}_{jm}, \hat{l}_{jn} )  \; .
\end{equation}
As a last step we can also compact the notation by defining 
\begin{equation}
	\label{eq:cal_R_def}
	\mathcal{R}_{ij}^{L/R} (k) \equiv  (2\pi k L_i) (2\pi k L_j) \, W(k L_i) \, W^*(k L_j)\,  \tilde{R}_{ij}^{L/R}(k) + h.c. \; ,
\end{equation}
to get
\begin{align}
	\label{eq:two_point_dfreq2_app2} 
	\left\langle s_i(t) s_j(t) \right\rangle  =  \int  \textrm{d} k \;  \left[ \mathcal{R}_{ij}^{L} (k) \; P_L (k) +\mathcal{R}_{ij}^{R} (k) \; P_R (k) \right] \; .
\end{align}
Notice that if $i,j$ are both LISA channels, we can trivially recover the response function $\mathcal{R}_{ij}(k) $ defined in~\cite{Flauger:2020qyi} as:
\begin{equation}
	\mathcal{R}_{ij}(k) = \mathcal{R}_{ij}^{L} (k) + \mathcal{R}_{ij}^{R} (k) = 16 (2 \pi k L )^2 \sin^2 \left( 2 \pi k L\right) \tilde{R}_{ij} (k)  \; , 
\end{equation}
where $\tilde{R}_{ij} (k) = \tilde{R}^L_{ij} (k) = \tilde{R}^R_{ij} (k) $.
 
 To conclude this section we stress that, by applying~\eqref{eq:Fourier} to~\cref{eq:two_point_dfreq2_app2} , we get
\begin{equation}
	\left\langle \tilde s_i(f) \tilde s_j^*(f') \right\rangle =  \frac{1}{2}   \left[ \mathcal{R}_{ij}^{L} (|f|) \; P_L (|f|) +\mathcal{R}_{ij}^{R} (|f|) \; P_R (|f|) \right] \, \delta_T(f - f') \, ,
\end{equation}
where~\footnote{In the limit of infinite observation time the $\delta_T$ can be replaced with a proper Dirac delta.} $\delta_T(f - f') \equiv T \text{sinc}(f-f')$. Notice that while $k$ is only positive, the $f$ and $f'$ appearing in this equations can be both positive and negative which is compensated by the $1/2$ prefactor. 

\subsection{Noise spectra and AET basis}
\label{app:noise}
The precise determination of the properties of the noise power spectra of the different channels is one of the main technical challenges of both the LISA and Taiji missions, and we will not treat this issue in detail here. The only current knowledge we have about the noise of space-based interferometers come from the Pathfinder experiment of LISA~\cite{Armano:2016bkm} and laboratory tests. Following~\cite{Flauger:2020qyi}, we will use the following expressions of the total power spectral density for the noise self-correlations of a single detector
\begin{align}
	\label{app:self_noise}
	N_{ii}(f, A, P ) =& 16 \sin^2(2 \pi f L) + \left\{ \left[ 3 + \cos(2 \pi f L) \right] P_{\rm acc}(f, A) + P_{\rm IMS}(f, P )\right\} \, ,\\
	\label{app:cross_noise}
	N_{ij}(f, A, P ) =& - 8 \sin^2(2 \pi f L)  \cos(2 \pi f L)+ \left[ 4 P_{\rm acc}(f, A) + P_{\rm IMS}(f, P )\right] \, ,
\end{align}
where $i, j \in {X,Y,Z}$ and $i \neq j$, $L$ denotes the detector arm-length, and
\begin{align}
	\label{eq:acc}
	P_{\rm acc}(f, A)  =& A^2 \, \frac{\mbox{fm}^2}{\mbox{s}^4 \, \mbox{Hz}} \left[ 1 + \left(\frac{0.4 \mbox{mHz}}{f}\right)^2 \right] \left[ 1 + \left(\frac{f}{8 \mbox{mHz}}\right)^4 \right] \left(\frac{1}{2 \pi f} \right)^4 \left(\frac{2 \pi f}{c} \right)^2 \, ,\\
		\label{eq:ims}
	P_{\rm IMS}(f, P )=&  P^2 \, \frac{\mbox{pm}^2}{\mbox{Hz}} \left[ 1 + \left(\frac{2 \mbox{mHz}}{f}\right)^4 \right] \left(\frac{2 \pi f}{c} \right)^2 \, ,
\end{align}
where $P_{\rm acc}$ and $P_{\rm IMS}$ denote, respectively, the ``acceleration" (acc) noise (associated with the random displacements of the masses caused, e.g., by local environmental disturbances), and the ``Interferometry Metrology System" (IMS) noise (which includes shot noise). For more technical details we refer the reader to~\cite{Flauger:2020qyi, Estabrook:2000ef, Tinto:2014lxa}. The noises of different detectors are assumed to be uncorrelated \emph{i.e.} $\langle \tilde{n}^i_j(f)  \tilde{n}^j_j(f')  \rangle = 0$ if $i$ is a LISA channel and $j$ is a Taiji channel~\footnote{As stated in the main text this assumes that both noise components are uncorrelated among the two detectors. This assumption must be tested for what concerns acceleration noise. }.

The derivation presented in~\cref{app:response} was carried out in terms of the so-called \text{XYZ} basis. We now introduce another commonly used basis of TDI channels, the so-called AET basis, which diagonalizes noise covariance matrix~\cite{Hogan:2001jn,Adams:2010vc} for the single detector~\footnote{Due to the detector symmetries, it is possible to show that also the signal is diagonal in the AET basis~\cite{Flauger:2020qyi}.}. In order to distinguish between the channels of different detectors, we use AET when referring to LISA and CDS when referring to the corresponding ``AET channels" of Taiji. Assuming that the noise spectra for all links are identical \emph{i.e.}
\begin{align}
	& N_{ \text{XX} } = N_{ \text{YY}  }=N_{ \text{ZZ} }\, , \\
	& N_{ \text{XY} } =N_{YZ} =N_{XZ} \, , 
\end{align}
it is trivial to show that the new basis
\begin{align}
	\tilde d_\text{A} =& \frac{1}{\sqrt 2} (\tilde d_\text{Z} - \tilde d_\text{X}) \, , \\
	\tilde d_\text{E} =& \frac{1}{\sqrt 6} (\tilde d_\text{X} - 2\tilde d_\text{Y} +\tilde d_\text{Z}  )\, ,\\
	\tilde d_\text{T} =& \frac{1}{\sqrt 3} (\tilde d_\text{X} + \tilde d_\text{Y} + \tilde d_\text{Z} ) \, ,
\end{align}
diagonalizes the LISA noise matrix (and an identical definition of CDS channels for Taiji diagonalizes the Taiji noise matrix).  With these definitions it is trivial to show that 
\begin{align}
	&N_{ \text{AA} } = N_{ \text{EE} } =N_{ \text{XX} }(f) - N_{ \text{XY} }(f) \, ,  \\
	&N_{\text{TT} } = N_{ \text{XX} }(f) + 2 N_{ \text{XY} }(f)  \, .
\end{align}
The complete expressions of $N_{ \text{AA} }$ and $N_{ \text{TT } }$ are given in~\cref{eq:AA_noise} and~\cref{eq:TT_noise}. It is possible to show that at low frequency $\mathcal{R}_{ \text{XY} } \simeq - \mathcal{R}_{ \text{XX} } / 2$ and as a consequence, in that regime we have $\mathcal{R}_{ \text{TT} } \ll \mathcal{R}_{ \text{AA} } $ so that T is signal orthogonal and it can be used to measure the noise parameters~\footnote{As shown in~\cite{Flauger:2020qyi} T mostly gives information on the $P$ parameter. }. 

\bibliographystyle{JHEP}
\bibliography{references.bib}
\end{document}